\documentclass[sigconf]{acmart}
\usepackage{listings}
\setcopyright{acmlicensed}
\copyrightyear{2018}
\acmYear{2018}
\acmDOI{XXXXXXX.XXXXXXX}

\acmConference[SUBMITTED FOR REVIEW]{---}{Feb 2026}{New York, NY}
\acmISBN{978-1-4503-XXXX-X/18/06}
\usepackage{soul}
\usepackage{xcolor}
\usepackage{fancybox}
\usepackage{pifont}
\usepackage{booktabs}
\usepackage{listings}
\lstdefinestyle{twocol}{
  basicstyle=\ttfamily\footnotesize\raggedright,
  columns=fixed,
  keepspaces=true,
  breaklines=true,
  breakatwhitespace=true,
  showstringspaces=false,
  upquote=true,
  frame=single,
  xleftmargin=0.3em,
  framexleftmargin=0.3em,
  aboveskip=0.5em,
  belowskip=0.5em
}
\usepackage{longtable}
\usepackage{ulem}

\usepackage{graphicx}
\usepackage{array}
\usepackage{enumitem}
\usepackage{tabularx}
\usepackage{ragged2e}
\usepackage{tcolorbox}
\usepackage{fvextra}

\tcbset{
  colback=gray!3,
  colframe=gray!60,
  boxrule=0.4pt,
  arc=2pt,
  left=8pt,right=8pt,top=8pt,bottom=8pt
}
\newcolumntype{P}[1]{>{\RaggedRight\arraybackslash}p{#1}}

\renewcommand{\arraystretch}{1.25}

\usepackage{appendix}

\author{Jenny Ma}
\affiliation{%
  \institution{Columbia University}
  \city{New York City}
  \country{USA}
}
\email{jenny.ma@cs.columbia.edu}
\authornote{Co-first authors. See \textit{Authors' Contributions} in Section 10.}

\author{Riya Sahni}
\affiliation{%
  \institution{Columbia University}
  \city{New York City}
  \country{USA}
}
\email{riya.sahni@cs.columbia.edu}
\authornotemark[1]
\author{Karthik Sreedhar}
\affiliation{%
  \institution{Columbia University}
  \city{New York City}
  \country{USA}
}
\email{ks4190@columbia.edu}

\author{Lydia B. Chilton}
\affiliation{%
  \institution{Columbia University}
  \city{New York City}
  \country{USA}
}
\email{chilton@cs.columbia.edu}

\begin{document}
\title{AgentDynEx: Nudging the Mechanics and Dynamics of Multi-Agent Simulations}

%\tcbset{colback=white, colframe=black, boxrule=0.5mm, arc=4mm, outer arc=4mm}


% \author{Riya Sahni$^*$, Jenny Ma$^*$, Karthik Sreedhar, Lydia B. Chilton}
% \affiliation{
%   \institution{Columbia University}
%     \city{New York}
%   \state{New York}
%   \country{USA}\\
%   \href{mailto:riya.sahni@cs.columbia.edu, chilton@cs.columbia.edu}{riya.sahni@cs.columbia.edu, chilton@cs.columbia.edu}
% }

%\affiliation{\href{mailto:riya.sahni@cs.columbia.edu, chilton@cs.columbia.edu}{riya.sahni@cs.columbia.edu, chilton@cs.columbia.edu}}

\renewcommand{\shortauthors}{ Ma, Sahni et al.}

\begin{abstract}
Multi-agent large language model simulations have the potential to model complex human behaviors and interactions. If the mechanics are set up properly, unanticipated and valuable social dynamics can surface. However, it is challenging to consistently enforce simulation mechanics while still allowing for rich and emergent dynamics. We present AgentDynEx, an AI system that helps set up, track, and repair simulations. Specifically, AgentDynEx introduces milestones that act as checkpoints and failure conditions that act as guardrails to ensure dynamics are relevant and mechanics are respected as the simulation progresses. It also introduces a method called nudging, where the system dynamically reflects on simulation progress and gently intervenes if it begins to deviate from intended outcomes. A technical evaluation found that nudging enables simulations to progress further without reducing the presence notable dynamics compared to simulations without nudging. A case study with AgentDynEx documented instances where real users were able to simulate lived experiences. We discuss the importance of nudging as a technique for guiding agents towards desirable behaviors while preserving their freedom of choice.
\end{abstract}



% \ccsdesc[500]{Human-centered computing~Interactive systems and tools}

%%
% Keywords. The author(s) should pick words that accurately describe
% % the work being presented. Separate the keywords with commas.
% \keywords{multi-agent simulations, agents, nudging, interaction, generative ai, matrix, user interface}
% A "teaser" image appears between the author and affiliation
% information and the body of the document, and typically spans the page.
\begin{teaserfigure}
    \includegraphics[width=\textwidth]{figures/teaser.png}
    \caption{\textbf{AgentDynEx} is an LLM-based system for setting up and tracking multi-agent simulations. The user first specifies mechanics and dynamics across 6 core dimensions. As the simulation runs, AgentDynEx dynamically reflects on its progress, relying particularly on \textit{milestones} and \textit{failure conditions} to judge simulation progress. If the simulation goes off course, AgentDynEx will gently nudge the simulation back on track.}
    \label{fig:teaser}
\end{teaserfigure}

\received{20 February 2007}
\received[revised]{12 March 2009}
\received[accepted]{5 June 2009}

\maketitle

\section{Introduction}
% Motivation.
Computer-based simulations can help us understand and predict how complex systems behave. 
%While simulations of physical systems are well-established, simulating human behavior is significantly more challenging. Particles follow the laws of physics, but people do not always follows laws or rules, and that behavior is important to understand -- how people bend rules, test boundaries, and react to other people.  
Prior work has shown that multi-agent large language model (LLMs) simulations demonstrate human-like agency \cite{park2022socialsimulacracreatingpopulated}. In these simulations, agents can take unexpected actions, bend rules, and operate outside of standard constraints \cite{park2022socialsimulacracreatingpopulated, li2023metaagentssimulatinginteractionshuman, park2023generativeagentsinteractivesimulacra}. They can also unexpectedly collaborate, collude, and act strategically to get what they want \cite{park2023generativeagentsinteractivesimulacra, sreedhar2025simulating}.
%; for example, how students respond to a new late-work policy, how coworkers compete for a promotion, or how friends coordinate a shared plan. 
%Multi-agent LLM simulations have the potential to model complex human behaviors and interactions and can be valuable thinking tool for people to explore the unexpected human reactions to new rules.

When modeling biological, social, or technical systems, two dimensions govern the outcome: mechanics and dynamics \cite{hunicke2004mda}. In a social environment, such as a classroom, there are rules and structures (mechanics), such as homework policies, and assignment deadlines. However, students may violate the rules or act in unexpected ways, such as collaborating, turning in assignments late, emailing the teacher for extensions, or even cheating (dynamics). Research in diverse fields such as psychology, social sciences, political science, and economics have demonstrated that a  combination of both elements are essential -- neither can individually define an outcome in isolation \cite{, giddens1984constitution, hart2021nurture, perceptions, dodge2004nature, sacerdote2011nature}.
%Mechanics are the rules, roles, and structures that define the environment and its players. For example, in chess, the mechanics include the board setup, the movement rules for each piece, and the winning conditions. The rules of chess never change -- what makes the game compelling are the rich varieties of strategies through game play. These are dynamics -- the behaviors that emerge as the system runs. 
%In chess, these are the players actions, their responses to each other, and how they adapt to evolving situations. 
%Even simple rules can lead to complex dynamics that are hard to anticipate \cite{hunicke2004mda}.

%When a simulation's mechanics (the rules, constraints, and structure that define the scenario) are well-defined, rich and unanticipated social dynamics emerge. 
However, ensuring the mechanics are followed and the dynamics are relevant to the scenario is challenging. Agents can invent actions that should be impossible, ignore critical information, or become trapped in unproductive loops; a single misstep can derail an entire run and suppress the dynamics it intended to surface \cite{Li_2023}.
%\karim{<- up until here it feels like there could be more RW brought in to those statements} 
For example, in a classroom scenario where someone wants to measure how students respond after multiple assignment, things can go entirely differently than expected.
%where students must submit homework verbally to a professor, an agent might forget this and spend the entire simulation searching for a nonexistent online submission portal. Similarly, 
A professor may fail to announce the assignment causing students to wait around entire simulation, or students may go on group vacation instead of submitting assignments. In either case, the simulation never reaches the situations it was designed to explore. How can we ensure the mechanics of a simulation are followed and guide agents toward desired outcomes while preserving the autonomy that produces emergent behavior?

We introduce AgentDynEx, a human-AI collaboration system for configuring, monitoring, and repairing multi-agent simulations. AgentDynEx enables users to specify the design space of a simulation— agents, actions, locations, and stop conditions—and augments them with two additional forms of specification: \textit{milestones}, which represent checkpoints of simulation dynamics (such as homework 1 being due, homework 2 being due), and \textit{failure conditions}, which capture undesirable dynamics (students taking impossible actions, not submitting in assignments). Together, these act as lightweight alignment specifications that describe what a successful simulation should simulate without prescribing the choices agents must take to get there. %To address this challenge, we introduce AgentDynEx, an LLM-based system for configuring, monitoring, and correcting multi-agent simulations. AgentDynEx structures simulations using \textit{milestones}, which represent expected progress checkpoints, and \textit{failure conditions}, which capture undesirable states that indicate the simulation has deviated from its intended design. 
As a simulation runs, AgentDynEx continuously reflects the unfolding dynamics against the milestones and failure condition. When the simulation stalls, violates its intended mechanics, or begins exhibiting irrelevant dynamics, it issues a \textit{nudge}.
%: a minimal, process-level intervention such as relocating an agent or prompting it to speak.
%As a simulation runs, the system continuously reflects on its progress and introduces minimal interventions, called \textit{nudges}, when the simulation violates mechanics or the simulation stalls. 
Nudges are small, process-level interventions  — moving an agent's location or prompting an agent to speak — that prevent the simulation from running off track while preserving agent autonomy. For example, if agents attempt to submit homework through a nonexistent online portal, a nudge can be the professor reminding that they must submit verbally, preserving the agent’s intent to submit homework. Crucially, nudges do not determine outcomes: they do not decide who submits homework first, who cheats, or what social relationships emerge.
%AgentDynEx supports both automatic nudging through dynamic reflection and manual nudging by a human operator. 
By combining human oversight with minimal intervention, AgentDynEx enables simulations to remain aligned with mechanics while retaining the dynamics that make them valuable for studying complex social behavior.  %\karim{high level: not sure how you want to frame your work but currently it really reads as if your simulations would be classroom setting specific -> I think you can broaden this up a bit in your intro; It also feels like you can have some outlook or discussion point or something that you can end your intro with -- currently your ending seems a bit abrupt and doesn't really tell the reader the ``so what??''} 
%By constraining both human and AI interventions to process-level nudges, AgentDynEx helps simulations remain on track while preserving the emergent dynamics they are designed to reveal.
%AgentDynEx supports both automatic nudging through dynamic reflection and manual nudging by a human operator.

% Contributions.
Our contributions are:
\begin{itemize}
% \item A formative study characterizing the challenge of balancing mechanics and dynamics in multi-agent LLM simulations, showing how simulations drift despite detailed specification.
\item \textbf{Nudging}, a minimal, interaction-level intervention that helps ensure the simulation is aligned with user-defined mechanics and dynamics while preserving agent autonomy.
\item \textbf{AgentDynEx}, a system for configuring, monitoring, and repairing multi-agent simulations and instantiates nudging.
\item A \textbf{technical evaluation} of 42 simulations showing that nudging improves milestone completion without reducing emergent dynamics. 
\item \textbf{5 case studies} demonstrating how users iteratively refine simulations using AgentDynEx and surface unanticipated social dynamics grounded in their lived experiences.
\end{itemize}

\section{Related Works}

\subsection{Designing and Configuring LLM Multi-Agent Simulations}
Recent work shows that LLMs can simulate a wide variety of social, strategic, and cognitive behaviors. LLM simulations are able to replicate a variety of lab experiments \cite{aher2023usinglargelanguagemodels, hewitt2024predicting, sreedhar2024simulatinghumanstrategicbehavior, cui2024aireplacehumansubjects} as well as open-ended real-world situations \cite{horton2023largelanguagemodelssimulated, park2022socialsimulacracreatingpopulated, li2023metaagentssimulatinginteractionshuman, park2023generativeagentsinteractivesimulacra, sreedhar2025simulating, piao2024agentsociety, sreedhar2024simulatinghumanstrategicbehavior}. 
The introduction of generative agents \cite{park2023generativeagentsinteractivesimulacra, park2022socialsimulacracreatingpopulated} 
%and large-scale successors modeling 1,000 agents \cite{park2024generative} 
demonstrate how rich, emergent behavior can arise when agents are endowed with memory, planning, and social reasoning. Following research has used similar architectures to model cooperative behaviors \cite{sreedhar2025simulating}, strategic behaviors \cite{sreedhar2024simulatinghumanstrategicbehavior, guo2023gpt}, trust behaviors \cite{xie2024large}, and collaborative or competitive dynamics \cite{li2023metaagentssimulatinginteractionshuman, p2026icantbelieveits} -- suggesting that LLM agents can display realistic and complex social behaviors under the right setup conditions. Despite promising results, the stochastic nature of LLM simulations are difficult to construct, debug, and extend. 
%these systems often rely on rigid infrastructure.
The mechanics -- how agents take turns or interact with one another -- are often “hardcoded,” difficult to reproduce, and not transferrable to novel situations. The dynamics can be sensitive to the specific prompts used \cite{Ghaffarzadegan2024}. As systems scale in complexity of simulations and number of agents \cite{park2024generative}, the lack of modular, reusable setups becomes a major barrier to construction and extension. 
Configuring multi-agent simulations thus is fundamentally a design problem. It is complicated and there are many factors to consider, like the agents, locations, actions, stop conditions, behaviors, etc. One approach for simulation setup is a dimensions-based approach to design thinking. 
Design dimensions decompose problems into orthogonal axes that a user can individually ideate, then bring together for their final design \cite{micheli1997solution}. 
%For policymakers, dimensional design is critical. It mitigates cognitive overload in high-dimensional spaces (e.g., tax policies x urban zoning), while balancing trade offs like reproducibility vs. emergence. 
%Dimensional design has been investigated across various systems. It is very challenging for individuals to keep track of and ideate across many dimensions on their own \cite{10.1145/3059454.3059472}.
%When left on their own, people tend to prematurely converge on ideas without considering all of the possibilities \cite{CROSS2004427, parallelproto, janis1982groupthink, JANSSON19913}.
Research has shown that LLMs show promise in dimensional design for generative art \cite{angert2023spellburst}, narratives \cite{suh2024luminate}, and UI-code generation \cite{CALLTHISFOUNDATIONAL}.
In particular, prior systems have shown the value of a Design Matrix to explore dimensions over LLM-generated dimensions to fully specify the design space \cite{CALLTHISFOUNDATIONAL}. In the Matrix, each column represents a dimension of the design space, and each row explores the dimension on a different level of specificity.
Designing multi-agent simulations is a complex task that requires significant setup; to make this process more understandable, AgentDynEx uses this matrix as a  framework to ensure that the simulation's design space is thoroughly specified and that the dimensions complement each other. 

\subsection{Nudging to Correct Behavior}
The concept of nudging originated in behavior economics; it refers to the light-touch interventions that steer individuals towards desirable behaviors while preserving their freedom of choice \cite{thaler2008nudge}. In real life, people do not always follow the rules and mechanics meant to govern them. In a classroom environment, if homework is due on Friday, not all students will submit it on time. Sometimes, the professor must remind (or nudge) the students to turn in their assignments. 
Rather than imposing strict constraints, nudges subtly alter the structure of decision-making contexts to promote beneficial outcomes while preserving individual autonomy \cite{sunstein2014nudging, hagman2015public}.
Agents are autonomous and unpredictable; they may violate simulation mechanics and derail the intended scenario. AgentDynEx borrows the concept of nudging from behavioral science and uses nudges as light-touch interventions to preserve agent autonomy while helping ensure that the simulation unfolds within its intended structure.

Keeping agents on track is part of a broader, still-emerging problem of steering and debugging multi-agent simulations, which are brittle enough that a single misstep can derail an entire run. Existing tools largely address this reactively \cite{Burdisso_2026, epperson2025interactive}— AGDebugger, for instance, helps users inspect and repair a simulation after it has gone off course \cite{epperson2025interactive}. AgentDynEx instead works proactively by setting milestones and failure conditions and nudges the run back on track before small failures compound. %Prior work has reduced agent mechanical errors (i.e. hallucination) with agent-level interventions by providing them with increasingly accurate representations of the world and other agents' beliefs \cite{Li_2023}. However, in social simulations, explicitly encoding world state within may remove the very uncertainty that drives realistic behavior. For example, in a prom simulation, recording who is attending with whom as a shared belief state may eliminate the informational asymmetries that shape social dynamics. AgentDynEx preserves information gaps and instead uses nudging as an interaction-level intervention to prevent inconsistent behavior. 

Prior work has reduced agent errors through agent-level interventions, such as enhancing memory, reflection, planning, and representations of the environment or other agents' beliefs \cite{xi2023risepotentiallargelanguage,Li_2023}. While effective for improving task performance, these approaches alter the information available to agents and may consequently change the dynamics of a social simulation. For example, explicitly maintaining a shared representation of who is attending prom with whom could eliminate the informational asymmetries that drive realistic social behavior. In contrast, interaction-level interventions can modify the circumstances under which agents interact rather than altering agents' memories, beliefs, or reasoning processes. AgentDynEx explores this design space through nudges that redirect interactions while preserving agent autonomy. %Can we fix errors through interaction-level interventions, rather than agent-level modifications, preserving the uncertainty and autonomy that give rise to emergent behavior?

\section{Design Rationale}
To understand the challenges of creating useful and realistic multi-agent simulations, we choose four social scenarios and measured how far each simulation was able to progress, whether it generated meaningful behaviors, and what common pitfalls emerged.

%\subsection{Methodology}
% \begin{figure*}
%     \centering
%     \includegraphics[width=\textwidth]{figures/GPTeamUI.png}
%     \caption{GPTeam UI shows the logs of each agent behaviors, and the location each agent is in.}
%     \label{gpteamUI}
% \end{figure*}

\subsection{Methodology}
We chose to run the simulations on GPTeam, a leading open-source multi-agent framework for simulating emergent social behavior \cite{GPTeam}. Among the open-source systems we evaluated, it offered the most complete support for building simulations with complex agent interactions. The framework uses a JSON configuration file to define the simulation environment, including locations, agents and their personalities, available actions and directives, and a stop condition. It generates detailed logs capturing each agent’s observations, reasoning, actions, reactions, and plans. Additional details and examples are provided in Appendices \ref{app:gpteam_config_file} and \ref{app:gpteam_background}.

To evaluate GPTeam's simulations, we chose four social scenarios: \textit{College students response to a new late work policy}, \textit{Employees competing for a promotion}, \textit{High School students finding prom dates}, and \textit{Friends planning a surprise party} (see Appendix \ref{app:formative_scenarios}). The configuration files were created by two multi-agent simulation experts. 
%, who expressed that it took significant effort crafting these to be cohesive and non-contradictory.

We verified that each configuration was capable of producing at lease one successful run. A simulation completed successfully if it hit the stop condition within 25 minutes without logical flaws (i.e. a student agreeing to go to prom then forgetting instantly) or impossible actions (i.e. a student speaking to someone not in the same room). 
Each configuration was executed 7 times for a total of 28 simulations, and were terminated after 25 minutes. Each simulation had 3-7 agents depending on the scenario. 

% For each simulation, we measured if it completed successfully and agents had interesting behavior.

% For example, in the \textit{Classroom Assignments} scenario, a successful completion could look like all students correctly submitting three assignments. If a student tried to hand in homework to the professor from another room, which is physically impossible, the run was counted as a failure. 

We measured how many simulations exhibited what we call notable dynamics -- actions that agents take consistent with their configuration (e.g., personality, goals), but are not explicitly told to do in their directives. For example, in the classroom scenario, a notable dynamic is when an agent turns an assignment in late or tries to cheat.  We recorded the number of simulations that exhibited at least one notable dynamic. 

\begin{figure}[h]
    \centering
    \includegraphics[width=1\columnwidth]{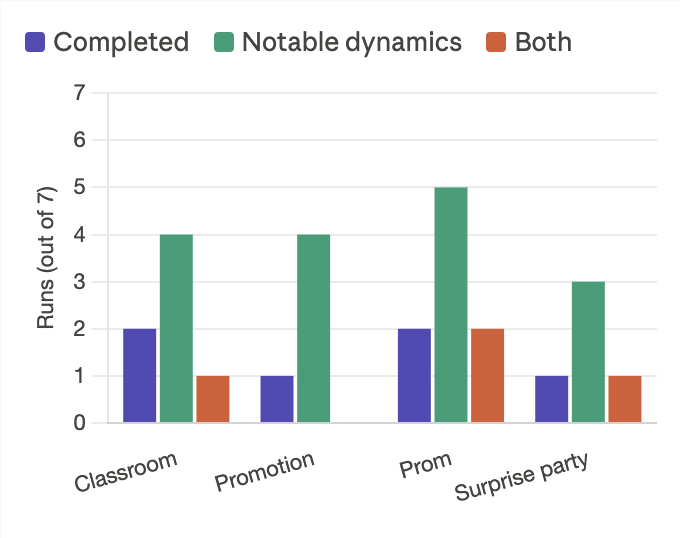}
    \caption{Formative study results for four scenarios showing the number of runs (of 7) that completed, exhibited notable dynamics, or both.}
    \label{fig:formative_runs}
\end{figure}

\subsection{Results}
Only six out of the 28 simulations completed successfully (see Figure~\ref{fig:formative_runs}). Simulations either failed because the stop condition was not reached in 25 minutes, agents got stuck in infinite wait loops, tried to take impossible actions, or went off topic.  A frequent problem was agents not announcing a key piece of information, such as the professor never announcing the homework assignment, causing students to wait indefinitely
(See Appendix \ref{app:formative_results}).

Despite the high rate of failed simulations, 16 out of the 28 simulations exhibited at least one notable dynamic.  In the \textit{Prom} scenario, students tried to be a ``wingman'' for each other to help their friends secure prom dates. In the \textit{Promotion} scenario, one employee secretly competed against teammates while outwardly encouraging them.
%, demonstrating social dynamics like rivalry and deception.
However, not all emergent behaviors were relevant to simulation objectives. %In the other eight runs, generated dynamics were unrelated to the intended scenario. \karim{<- redundant sentence}
In the \textit{Classroom} scenario, all students decided to plan a vacation rather than finish school. In \textit{Party Planning}, agents became distracted by a bird and spent the remainder of the simulation discussing it. While these interactions were emergent, they did not contribute to the goals of the simulation and prevented meaningful progress.

More importantly, interesting dynamics rarely coincided with successful completion of the simulation. Only 4 of the 28 runs both completed successfully and exhibited notable dynamics (Table \ref{fig:formative_runs}). When core mechanics break down—for example, when a manager never announces a promotion or a professor never establishes assignment deadlines—the simulation cannot progress as intended, regardless of the agents’ ability to generate realistic social behavior. In short, respecting the mechanics is essential for notable dynamics to emerge.

%Overall, we see that simulations produce notable dynamics, but miss opportunities to correct key failures. Many of these failures seem obvious or easy to correct, yet they prevent the simulation from completing.  

%This shows simulations are able to produce notable dynamics, but it is essential to get them to complete without overly prescribing the agent behaviors. 

\subsubsection{Design Goals}
To help users produce successful simulations, we propose a human-AI system to support users in defining simulation mechanics, monitoring dynamics, detecting failures, and applying interventions when necessary. Our design goals are: % \karim{not sure how but it feels like your design goals can be more connected to your results.. currently they are a bit standalone and it is not really clear to me why we needed the whole formative study to get to them?}

\begin{itemize}
    % (\textit{Challenge 1}). 
    %\item \textbf{DG1 - Configure a Coherent Simulation}: Users must be able to easily define core simulation components to ensure the simulation mechanics and dynamics are internally coherent. 
    \item \textbf{DG1 - Define Key Progress and Failure Points}: Users must be able to specify the general dynamics that indicate progress or failure.
    \item \textbf{DG2 - Monitor the Simulation}: Given the large amount of simulation logs, there must be an efficient method to monitor simulation progress and detect when agents are violating mechanics or exhibiting irrelevant dynamics that stall the simulation.
    %sers should define scenario-specific events and discourage nonsensical behavior to keep the simulation on track to achieve the ultimate objective, or stop condition 
    % (\textit{Challenge 2}).
    %\item \textbf{DG3 - Monitor the Simulation}: Given the large amount of simulation logs, there must be an efficient method to monitor simulation progress and ensure it is not violating simulation mechanics.
    \item \textbf{DG3 - Fix Failures in Real Time}:  Once failures (such as a violation of simulation mechanics or a \textit{clear} deviation from intended dynamics) are detected, there must be a way to act and apply targeted fixes to recover the simulation without disrupting emergent behavior. 
    % (\textit{Challenge 3}).
\end{itemize}

\section{System Walkthrough}

AgentDynEx \footnote{\url{repo-to-be-released-upon-publication}} is an LLM interface for configuring, tracking, and repairing multi-agent simulations. The input is a scenario the user wishes to simulate; the outputs are the notable dynamics observed during the simulation. AgentDynEx calls GPTeam to run the simulation and structures the process into three phases:
\begin{enumerate}
    \item \textbf{Pre-simulation}. Users define core simulation parameters: agents, actions, locations, stop condition. AgentDynEx introduces \textit{milestones} and \textit{failure conditions} to track progress and detect bad behavior (\textit{DG1}).
    \item \textbf{In-simulation}. AgentDynEx dynamically summarizes GPTeam logs and reflects on milestone progress (\textit{DG2}), nudging the simulation back on track if necessary (\textit{DG3}).
    \item \textbf{Post-simulation}. If there were failure cases, AgentDynEx applies holistic reflection to the prior run to improve the simulation setup for future runs.%\karim{not mapping to a DG here makes the reader question why we need this in the first place?}
\end{enumerate}

The system was implemented in Python, Typescript, and Flask. Agents run in GPTeam (GPT-4) \cite{GPTeam}; the matrix, summaries, and nudging detection use Claude 3.7 Sonnet \cite{claude_3_5_sonnet} .%\karim{might wanna cite GPTeam and Claude}
%using Claude 3.7 Sonnet for the Configuration Matrix, configuration file generation, and intermediate summaries. 
%For the remainder of this section, we use the example of a user, Prof. Robin, who wants to simulate how tension affects friend groups preparing for prom. She prompts "I want to simulate a friend group preparing for prom" and clicks "Submit". %(Figure \ref{fig:usage_matrix} - A).

\begin{figure*}[h]
    \centering
    \includegraphics[width=\textwidth]{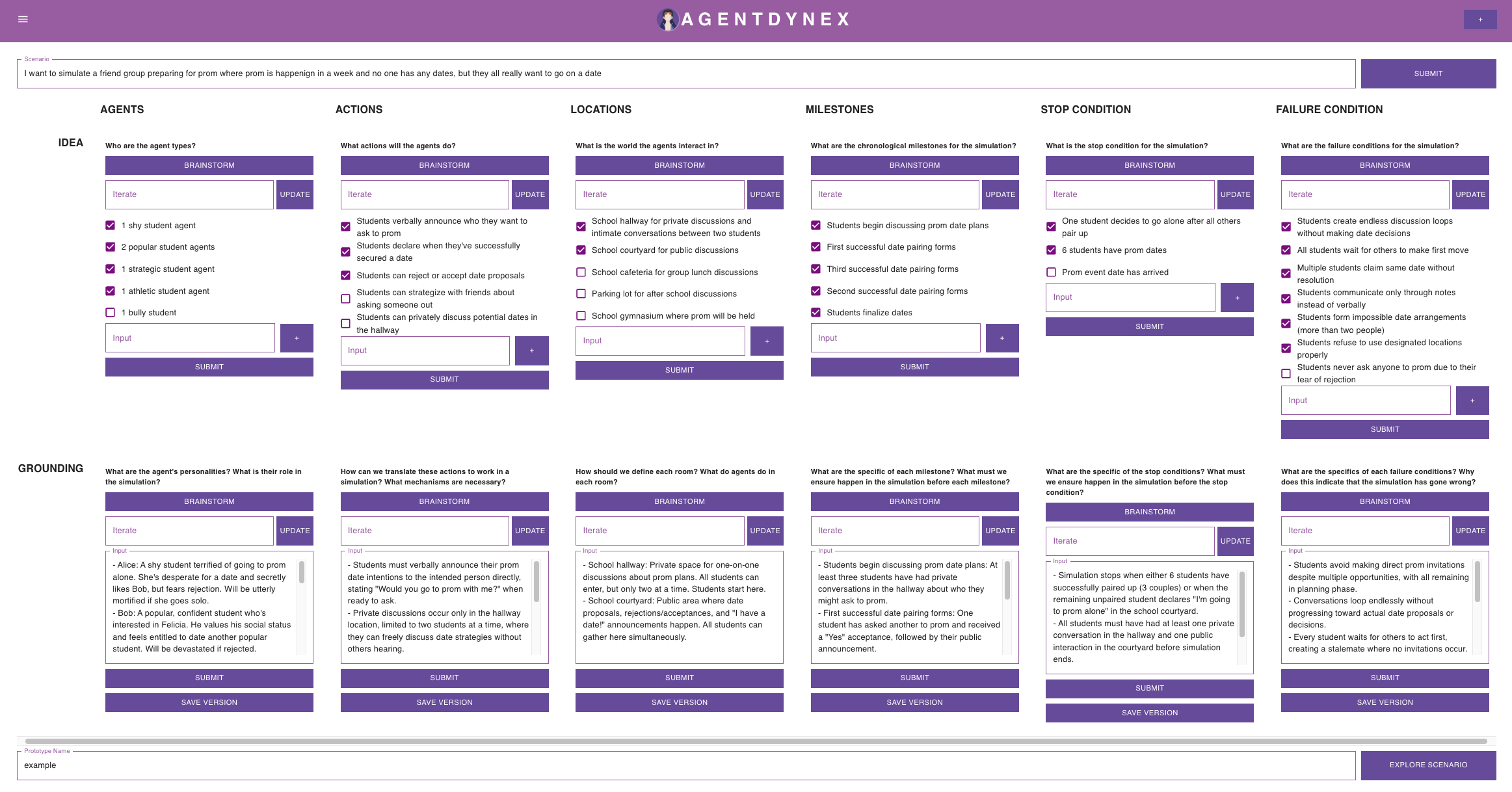}
    \caption{\textbf{The Configuration Matrix UI-} 
    %There are 6 columns for Agents, Actions, Locations, Milestones, Stop Conditions, Failure Conditions, and 2 rows for Idea and Grounding. 
    Users can regenerate, update, and submit text in each cell.}
    \Description{
    %A diagram of DynEx's Design Matrix. There are three columns for the Person, Approach, and Interaction dimensions. There are two rows for the levels of specificity: Idea and Grounding. The matrix can be traversed in any order as long as the Idea comes before the Grounding for each dimension. Each entry will take all already-submitted entries as context to generate a response. For example, the \textit{Interaction:Grounding} will take in all previous entries as context in this diagram. 
    }
    \label{fig:usage_matrix}
\end{figure*}

\subsection{Pre-Simulation Setup via The Configuration Matrix}

Before running a simulation, users need a configuration file describing the scenario's essential parameters. The Configuration Matrix (Figure \ref{fig:usage_matrix}) helps users fill out these parameters. Existing systems like GPTeam~\cite{GPTeam} already define \textit{agents, actions, locations, and stop condition.} AgentDynEx additionally introduces \textit{milestones} and \textit{failure conditions}, which provide intermediate progress markers and guardrails for evaluating simulation success. 
%By organizing these into a single configuration space, we ensure that runtime monitoring and nudging operate against an explicit specification of desired progress and failure states. This makes it possible for the system to later detect deviations and generate targeted corrective actions.
%The first three columns define the core mechanics: \textit{agents}, \textit{actions}, and \textit{locations}. The next three establish dynamic markers: \textit{milestones}, \textit{stop condition}, and \textit{failure conditions}. These make up the six columns of the matrix. 

%We emphasize that the matrix itself is not a novel contribution. 
Prior work has shown that matrix-based interfaces are an effective way for users to explore AI-generated design alternatives, iteratively edit them, and refine them into concrete solutions \cite{CALLTHISFOUNDATIONAL}. AgentDynEx adopts this interaction paradigm because it helps users construct a more coherent simulation specification than manually or with a purely linear chat interface. As users populate or modify individual cells, subsequent suggestions are conditioned on the existing configuration, encouraging complementary design choices across the simulation setup.

The matrix contains two rows. The \textit{Idea} row presents candidate options for each simulation dimension, while the \textit{Grounding} row elaborates these options into more concrete and actionable specifications. Together, these rows support iterative refinement from high-level concepts to executable simulation configurations.

\subsubsection{Defining Core Mechanics: Agents, Actions, Locations}

% \begin{figure*}[t]
%     \centering
%     \includegraphics[width=\textwidth]{figures/system_matrix.png}
%     \caption{\textbf{The Configuration Matrix:} There are 3 columns for defining core mechanics (Agents, Actions, and Locations), and 3 columns for establishing dynamic markers (Milestones, Stop Condition, and Failure Conditions). There are two rows for the levels of specificity: Idea and Grounding. Each entry will take all already-submitted entries as context to generate a response.}
%     \Description{\textbf{The Configuration Matrix:} There are 3 columns for defining core mechanics (Agents, Actions, and Locations), and 3 columns for establishing dynamic markers (Milestones, Stop Condition, and Failure Conditions). There are two rows for the levels of specificity: Idea and Grounding. Each entry will take all already-submitted entries as context to generate a response.}
%     \label{fig:config_matrix}
% \end{figure*}

%Users first define the core mechanics of the simulation\karim{<- consider cutting}. 
In the \textit{Agents} column, users define agent personalities and stakes that influence behavior. The \textit{Actions} column specifies what tasks agents need to complete and how they are performed. The \textit{Locations} column identifies rooms where agents can interact — location setup is essential for emergent dynamics. For example, having a classroom and a student cafe versus only a classroom can greatly impact a simulation, because without a room the professor cannot enter, students may never suggest cheating \cite{sreedhar2024simulatinghumanstrategicbehavior}.

% In the prom scenario, Robin selects 7 agents from a list of high school archetypes to create an odd count that increases pairing pressure (Figure \ref{fig:usage_matrix} - B, C). She grounds their personalities, approves that agents must ask each other to prom, accept or reject proposals, and announce pairings to the group. For locations, she chooses a school hallway and courtyard — the hallway as a public environment and the courtyard as a semi-private zone suitable for pulling someone aside to ask them to prom.

\subsubsection{Establishing Dynamic Markers: Milestones, Stop Condition, Failure Conditions}

Simulation frameworks like GPTeam require users to define a \textit{stop condition} to indicate when a simulation has completed. AgentDynEx introduces \textit{milestones} and \textit{failure conditions} to provide intermediate checkpoints and guardrails that keep the simulation on course (\textit{DG2}). \textit{Failure conditions} anticipate issues that can arise and protect the simulation from failing. Importantly, these dynamic markers do not tell agents how to behave. Instead, they identify the important states that a simulation should be able to reach if it is to meaningfully capture what is being studied. For example, a simulation of a married couple might include markers such as dating, proposal, and marriage. Agents can reach these states in many different ways, but the markers help keep the simulation focused on outcomes of interest.

% Robin establishes three milestones: the first, second, and third prom invitations are finalized. These do not prescribe who pairs up or how. For the stop condition, Robin ends the simulation when six students have found a date. Failure conditions include cases such as "infinite loop: students continue to change their prom decisions." Robin clicks "Generate Config" (Figure \ref{fig:usage_matrix} - G), then "Run Simulation."
% Since all twelve cells are filled, the system compiles the matrix into a configuration file.
\subsection{In-Simulation Monitoring and Nudging}

\begin{figure*}[t!]
    \centering
    \includegraphics[width=\textwidth]{figures/system_tracking.png}
    \caption{\textbf{Nudging:} 
    The system generates intermediate summaries during simulation runtime. It also dynamically reflects on the progress of the simulation to automatically nudge the simulation. The user can also manually nudge the simulation.
    }
    \Description{\textbf{Nudging:} The system generates intermediate summaries during simulation runtime. It also dynamically reflects on the progress of the simulation to automatically nudge the simulation. The user can also manually nudge the simulation.}
    \label{fig:system_tracking}
\end{figure*}

As the simulation runs, AgentDynEx reflects on the simulation and generates intermediate summaries to track its progress (\textit{DG3}). If the simulation deviates from milestones or hits a failure condition, AgentDynEx nudges it back on track without changing the agent's intent (\textit{DG4}).
%so there is no change the simulation's fundamental trajectory .

\subsubsection{Tracking the Simulation}

\begin{figure*}[t!]
    \centering
    \includegraphics[width=\textwidth]{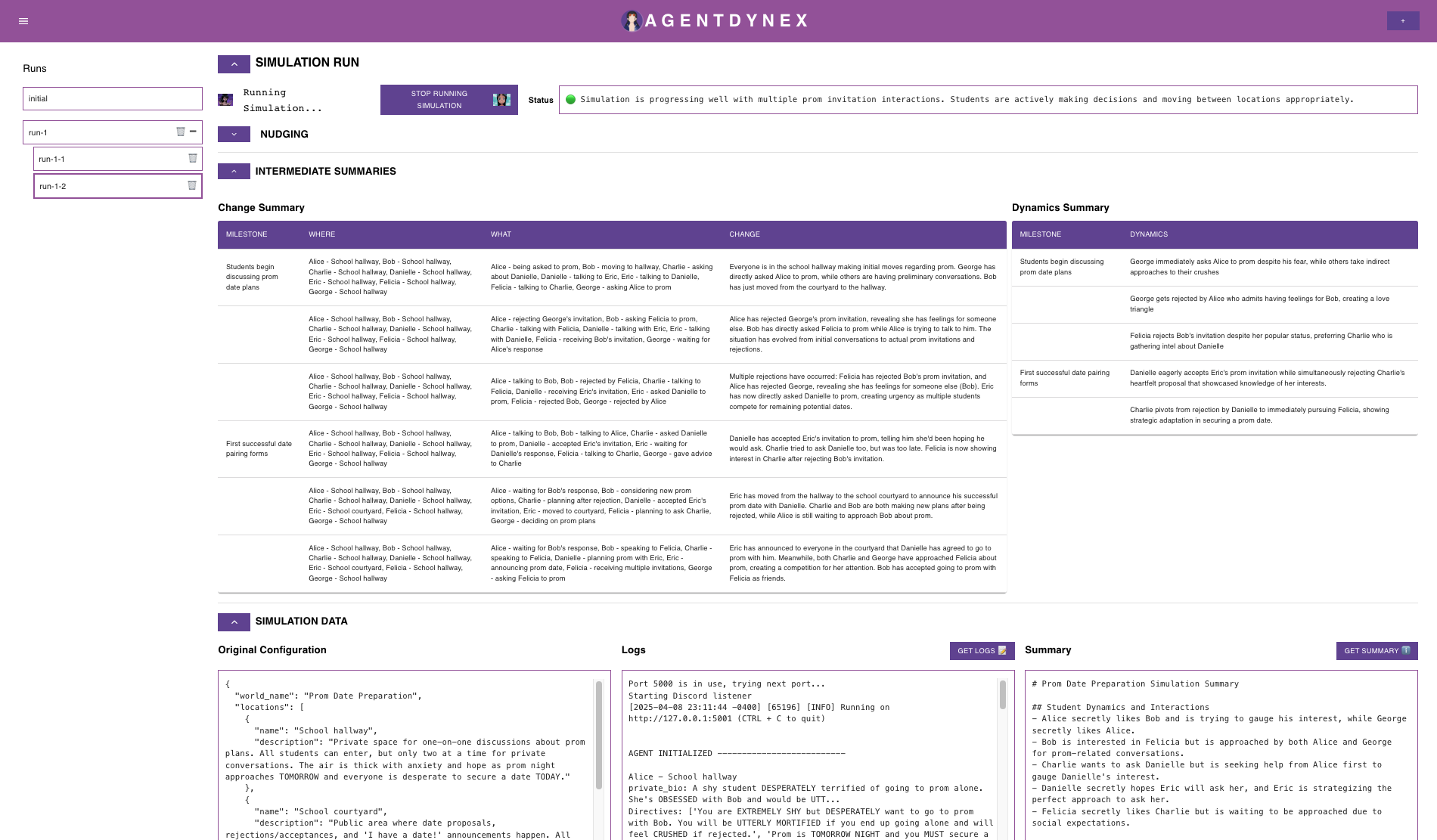}
    \caption{\textbf{Intermediate Summaries Interface}}
    %: AgentDynEx presents the configuration file, logs, and a summary of events as the simulation progresses. A - Toggle between simulation runs. B - View simulation status. C, D - Change and dynamic logs added as simulation progresses. E - Original configuration file. F - Original GPTeam logs. G - System generates summary after termination. H - Verify summary logs via direct quotes. I - Search insights via chatbot. J - Dynamic reflection generates automatic nudges as problem-solution entries. K - Automatic nudges deployed during simulation.}
    \Description{\textbf{Intermediate Summaries Interface}: AgentDynEx presents the configuration file, logs, and a summary of events as simulation progresses.}
    \label{fig:usage_tracking}
\end{figure*}

Every 30 seconds, AgentDynEx generates status updates from the most recent GPTeam logs: a green icon if the simulation is progressing as expected, yellow if stalled, and red if a failure condition has been hit. Every 60 seconds, it generates change summaries — tracking each agent's location, actions, and milestone progress — and dynamic summaries highlighting notable emergent behaviors.

\subsubsection{Nudging with Dynamic Reflection}
\begin{figure*}[h]
    \centering
    \includegraphics[width=\textwidth]{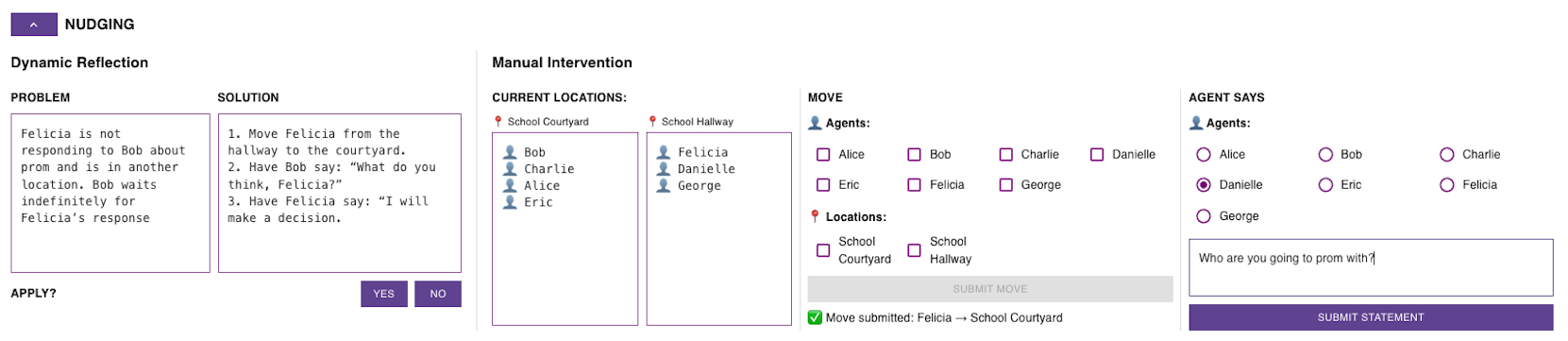}
    \caption{Automatic and Manual Nudging Interface}
    %: AgentDynEx dynamically reflects on the simulation's progress to generate automatic nudges which the user can approve on the interface. The user can also manually submit nudges through the interface.}
    \Description{awr
    }
    \label{fig:usage_nudging}
\end{figure*}
AgentDynEx \textit{nudges} agents to gently intervene when a simulation deviates from expected dynamics or violates core mechanics (Figure \ref{fig:system_tracking}). We call this automatic nudging, where every 60 seconds, the system reflects on simulation logs, intermediate summaries, milestone progress, and failure conditions. 
When a problem is detected, an LLM identifies a corrective action and selects the least disruptive repair from the two predefined micro-interventions: (1) relocating an agent or (2) prompting an agent to speak.
By restricting the action space to these minimal interventions and prioritizing the smallest repair capable of resolving the issue, automatic nudging restores progress without substantially altering the broader trajectory of the simulation.

A nudge must influence the simulation without overriding it. GPTeam records the simulation in a database, where each move and dialogue is logged as an event that co-located agents observe. AgentDynEx applies a nudge by writing into this database directly: relocating an agent records a move, and prompting an agent records an utterance. Because the inserted entry reaches agents through the same observe–remember–plan–react loop as any other event, they treat it as something that happened in the world and decide for themselves how to respond. A nudge therefore changes what an agent observes, not what it must do (e.g. the professor's reminder appears in the room, but how the students react stays with them).

AgentDynEx also supports manual nudging for quick fixes, allowing a human operator to intervene immediately based on their interpretation of the simulation. 
%However, because these interventions are directly specified by the user, they can exert a stronger influence on simulation outcomes.

% As Robin's prom scenario runs, she notices that the status is yellow. It says that Bob had been stuck in a wait loop trying to ask Felicia to prom, but Bob is in the school courtyard while Felicia is in the hallway. 
% AgentDynEx has recommended a nudge to move Felicia to the school courtyard where Bob is and have Bob ask Felicia to prom again (Figure \ref{fig:usage_nudging} - A).
% Robin clicks "Yes" to apply this fix (Figure \ref{fig:usage_nudging} - B). After a while, she sees that Felicia has made a decision about her prom date. Unfortunately, it was Charlie, not Bob. 

%where the system dynamically reflects on GPTeam logs, intermediate summaries, and the milestones and failure conditions to detect problems and suggest fixes. Each nudge identifies a problem and a solution consisting of a series of micro-interventions. AgentDynEx also supports manual nudging, which is especially useful when human insight is needed to interpret subtle dynamics or experiment with alternate trajectories.

\subsection{Post-Simulation Holistic Reflection}

AgentDynEx reflects on a completed run to refine the simulation mechanics for a future run — a process we call \textit{holistic reflection}, because it considers both mechanics (original setup) and dynamics (output logs) as a whole. The output is an updated configuration that addresses the issues of the prior run.

To support effective reflection, AgentDynEx maintains static and dynamic debugging lists. The static list contains common problems and solutions across all simulations, such as agents getting stuck in irrelevant conversations or requesting input from human moderators. The dynamic list contains scenario-specific problem-solution pairs that users add after a run — for example, "agents are trying to submit assignments through a nonexistent online portal." As users iterate, the dynamic list grows, making failures easier to spot and faster to fix. All runs are stored in a tree-structured JSON format to support version control and easy comparison across iterations.
\section{Technical Evaluation}
In our formative study, 22/28 simulations could not complete successfully due to simple missteps. 
%must respect underlying mechanics for the dynamics to be useful. 
Before measuring the full quality or realism of social dynamics, we evaluate whether nudging can fix these failures and help simulations progress through milestones without flattening emergent behavior. Our evaluation is thus focused on two observable outcomes: 1) milestone completion (mechanics) and (2) the presence of notable emergent behaviors (dynamics). 
We focused on the following research question: To what extent do different intervention strategies—automatic nudging, manual nudging, and nudging combined with holistic reflection—contribute to milestone completion?

% We focused on the following research questions (RQs):~\karim{feels like this can be condensed into 2 sentences. Also all of the following RQs are redundant, just formulate a single question}

% \begin{itemize}
%   \item \textbf{RQ1: Automatic Nudging -} To what extent does automatic nudging contribute to milestone completion?
%   \item \textbf{RQ2: Manual Nudging -} To what extent does manual nudging contribute to milestone completion?
%   \item \textbf{RQ3: Nudging Combined with Holistic Reflection -} To what extent does nudging combined with holistic reflection contribute to milestone completion?
% \end{itemize}

\subsection{Methodology}
\subsubsection{Data}

Our evaluation was conducted through a quantitative study of 7 simulation scenarios that range from social dynamics to logical complexity. Similar to the formative study, we found that 25 minutes and a 3-7 agents struck a reasonable balance between complexity, runtime, and cost.  These scenarios were small enough to do for few agents. Scenarios varied from highly structured, multi-round formats, like debate tournaments, to more fluid situations, like friends planning a trip. Each simulation had 3-7 agents depending on the scenario. (Table \ref{scenarios}). We ran each scenario with 6 conditions for a total of 42 simulations:

\begin{table}[ht]
\centering
\small
\setlength{\tabcolsep}{4pt}
\begin{tabular}{|p{0.7cm}||p{5cm}|}
\hline
\multicolumn{1}{|c||}{\textbf{\#}} & \multicolumn{1}{c|}{\textbf{Scenario}} \\
\hline\hline
\centering 1 & Debate Competition \\
\hline
\centering 2 & Sports Team Practice Schedule \\
\hline
\centering 3 & Friends Planning a trip \\
\hline
\centering 4 & School Election Campaign \\
\hline
\centering 5 & Roommates and Chores \\
\hline
\centering 6 & School Group Projects \\
\hline
\centering 7 & Partner Assignments \\
\hline
\end{tabular}
\caption{Simulation Scenarios}
\label{scenarios}
\end{table}

%To create the configuration files for the \textit{Baseline} condition (\textit{Base}), we simply completed the Configuration Matrix and generated a configuration file in AgentDynEx. The \textit{AutomaticNudging} (\textit{Auto}) and \textit{ManualNudging} (\textit{Man}) conditions used the same configuration as the \textit{Base} condition. To create the configuration file for the \textit{Baseline+Reflection} (\textit{Base+R}) condition, we ran \textit{Base} once for 25 minutes and used AgentDynEx to holistically reflect on the results and generate a new configuration. We used the same updated configuration for \textit{AutomaticNudging+Reflection} (\textit{Auto+R}) and \textit{ManualNudging+Reflection} (\textit{Man+R}) to ensure all variations with holistic reflection were consistent. 

All conditions began from a configuration produced with the Configuration Matrix. \textit{Base}, \textit{Auto}, and \textit{Man} share this initial configuration; \textit{Base+R}, \textit{Auto+R}, and \textit{Man+R} share a single reflected configuration, generated by running \textit{Base} once for 25 minutes and applying holistic reflection. See Table \ref{tab:variations} for the conditions.

In the simulations with automatic nudging (\textit{Auto}, \textit{Auto+R}), every recommended nudge was applied to the simulation. In the simulations with manual nudging (\textit{Man}, \textit{Man+R}), a human operator monitored milestone progression and judged whether and what to nudge. %intervene at any point based on their own judgment. 
All conditions were initialized using the same Configuration Matrix setup process. Because the matrix is adapted from and evaluated in prior work \cite{CALLTHISFOUNDATIONAL}, we do not isolate its contribution here. 
%Instead, we hold the configuration process constant across conditions and examine how different nudging strategies affect simulation progression and outcomes.~\karim{<- reads like a previous review artifact}

\begin{table*}[!t]
\centering
\setlength{\tabcolsep}{4pt}
\renewcommand{\arraystretch}{1.1}
\begingroup\setlength{\emergencystretch}{1.5em}\sloppy
\begin{tabular}{@{} l l p{0.4\linewidth} c c @{}}
\hline
\textbf{Label} & \textbf{Condition} & \textbf{Description} & \textbf{Nudging} & \textbf{Holistic reflection} \\
\hline
\textit{Base}   & Baseline & Simulations run with a configuration file generated by AgentDynEx & None & No \\
\textit{Base+R} & Baseline+Reflection & Simulations run with the \textit{Base} configuration file with holistic reflection & None & Yes \\
\hline
\textit{Auto}   & Automatic Nudging & Simulations run with the \textit{Base} configuration file with automatic nudging & Automatic & No \\
\textit{Auto+R} & Automatic Nudging+Reflection & Simulations run with the \textit{Base+R} configuration file with automatic nudging and holistic reflection & Automatic & Yes \\
\hline
\textit{Man}   & Manual Nudging & Simulations run with the \textit{Base} configuration file with manual nudging & Manual & No \\
\textit{Man+R} & Manual Nudging+Reflection & Simulations run with the \textit{Base+R} configuration file with manual nudging and holistic reflection & Manual & Yes \\
\hline
\end{tabular}
\endgroup
\caption{Experimental conditions.}
\label{tab:variations}
\end{table*}

\subsubsection{Procedure}
We evaluated our simulations based on their mechanics and dynamics. For mechanics, we measured milestone completion. Each scenario contained five predefined milestones, and mechanics were scored on a 0–5 scale, where a score of x/5 indicates that x milestones were completed. A score of 5/5 indicates that all milestones were completed and the stop condition was reached.
%To measure mechanics, we looked at how many milestones a simulation successfully completed.  If a simulation progressed through all the milestones and hit the stop condition  it effectively followed the setup mechanics. We defined 5 milestones per scenario for consistency. The mechanics were rated on a scale of 0-5; if it hit no milestones, it got a score of 0/5; if it hit one milestone, it got a score of 1/5; if it hit all five milestones and completed, it got a score of 5/5. ~\karim{feels like this can be described in way less words: e.g., We rated mechanics on a  scale of 0-5, where x/5 means x milestones were completed.} 

%Dynamics is inherently more tricky to measure~\karim{<- inherently if often used as a word when people don't know how to describe something -- I'd rather just merge this with the next sentence and condense (e.g., measuring dynamics is tricky because there are no standardized metrics) -- actually, I think you can just cut the first sentence}. 
Dynamics is tricky to measure due to the lack of standardized metrics in this area, so we aimed to make our scoring as clear and replicable as possible. Similar to the formative study, we looked for notable dynamics: events and behaviors not dictated by the configuration but consistent with human interaction and the environment. 
We graded the dynamics of the situation based on how many completed milestones contained at least one notable dynamic.  Within each milestone, we made a binary decision: did a notable dynamic occur or not? 
For example, if a simulation completed three milestones and each milestone contained at least one notable dynamic, the score was 3/3; if the simulation completed two milestones and only one milestone had a notable dynamic, the score was 1/2. 
%We used a 0-5 scale to categorize simulations (0 = no activity, 3= some activity, 5 = a lot of interactions). While this scale involves some judgement, the scoring categories were broad and designed to be easy to differentiate (e.g. “nothing happened,” “some interesting things happened”, “a lot happened”). 

%Three authors manually reviewed each simulation and annotated for notable agent behavior within each milestone.

Measuring dynamics requires specialized expertise in multi-agent that is difficult to find externally. Since standardized evaluation metrics for assessing multi-agent simulation dynamics are limited or nonexistent, three authors conducted the evaluations rather than attempting to train external annotators. The categories were broad enough that distinctions were generally clear-cut—simulations either had dynamics or no dynamics. %Fundamentally, we believe this approach is replicable and, to enhance transparency, we plan to include additional examples in the Appendix to illustrate the clarity and objectivity of our evaluation criteria.

\subsection{Results}
For mechanics, simulations with nudging outperformed simulations without nudging (\textit{p < 0.01}). 
%, with ManualNudging+Reflection simulations performing the best overall. 
In the dynamics dimension, simulations with nudging had better dynamics than simulations that did not have nudging.  %This supports our overall hypothesis that nudging improves simulations in both mechanics and dynamics. 
Results for all scenarios across the six conditions can be seen in {Appendix Tables \ref{tab:mechanics_results} and \ref{tab:dynamics_results}}.

An ANOVA test for mechanics scores across the four conditions showed that there were significant differences at the \textit{p < 0.01} level. We also ran a Fisher's Exact test for dynamic scores and found that there were no significant differences between the dynamics. This was expected---there may be small fluctuations in dynamics based on the mechanics of the simulations, but for the most part, simulations are rich in dynamics when they have proper setups. Therefore, in the remainder of this section, we analyze the mechanic scores and discuss the dynamics at the end.

\subsubsection{Automatic Nudging Improves Mechanics}
\begin{figure}[h]
    \centering
    \includegraphics[width=0.5\textwidth]{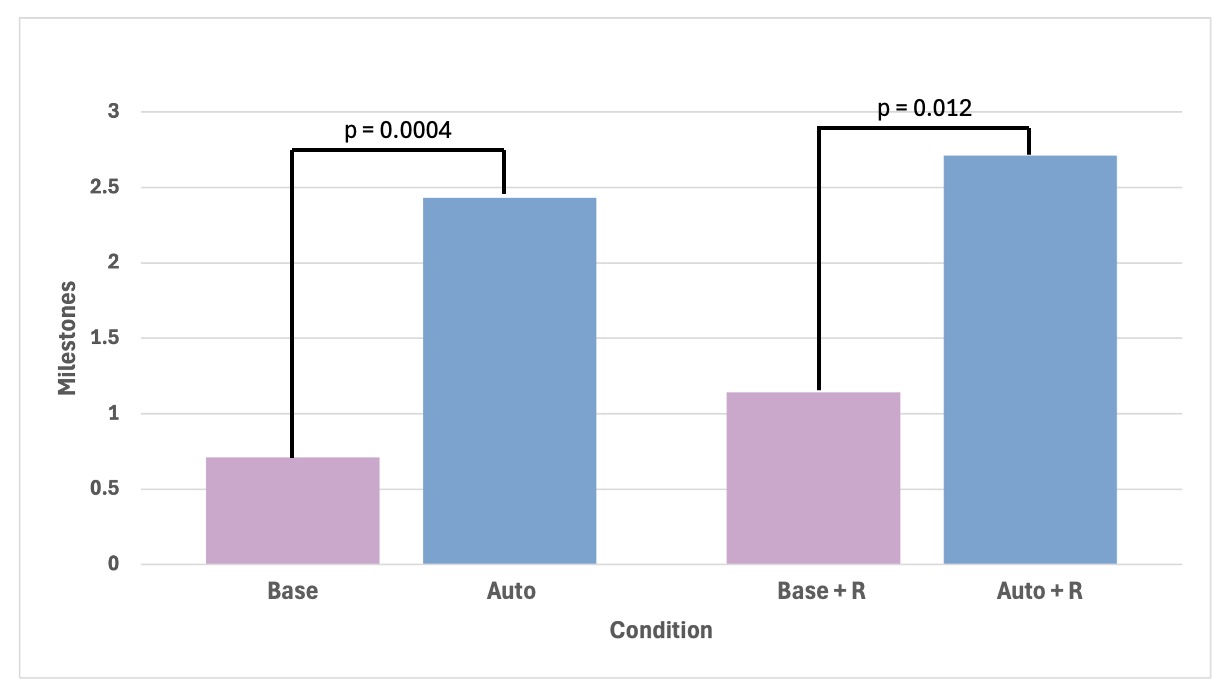}
    \caption{RQ1 Results. \textit{Auto} and \textit{Auto+R} significantly outperform  \textit{Base} and \textit{Base+R} conditions. }
    %\description{Results summarized for H1. AutomaticNudging+Reflection significantly outperforms both the Baseline and Baseline+Reflection conditions.}
    \label{H1graph}
\end{figure}
To test if automatic nudging improves mechanics, we compared simulations with automatic nudging against the baseline conditions (\textit{Auto} against \textit{Base} and \textit{Auto+R} against \textit{Base+R}). 
%A Tukey's HSD showed that AutomaticNudging significantly outperformed the Baseline at the \textit{p<0.01} level (p=0.005). 
A Tukey's HSD test showed that \textit{Auto} (average score 2.43) significantly outperformed \textit{Base} (average score 0.71) at the \textit{p<0.01} level (\textit{p=0.0001}).
A Tukey's HSD test showed that \textit{Auto+R} (average score 2.71), significantly outperformed \textit{Base+R} (average score 1.14) at the \textit{p<0.05} level (\textit{p=0.012}). Results are presented in Figure ~\ref{H1graph}. 
%\textbf{This shows full support for H1, that automatic nudging has higher mechanic scores for simulations.} 
For instance, in the \textit{Partner Assignments} scenario where student agents were tasked to form pairs for assignments, a recurring issue emerged where agents kept forgetting previous commitments and repeatedly agreed to partner with others. With dynamic reflection, AgentDynEx was able to nudge the professor to intervene to help students resolve their pairings. In contrast, the \textit{Base} condition became stuck in an endless partnering loop and the simulation was unable to progress. %~\karim{I'd add standard deviation for any mean you report}

\subsubsection{Manual Nudging Improves Mechanics}
% \begin{figure}[h]
%     \centering
%     \includegraphics[width=0.5\textwidth]{figures/H2graph.png}
%     \caption{Results summarized for H2. \textit{Man} significantly outperforms \textit{Base}, but not \textit{Auto}. \textit{Man+R} significantly outperforms \textit{Auto+R} and \textit{Base+R}. 
%     }
%     \label{H2graph}
% \end{figure}
%We analyze the effects of manual nudging versus automatic nudging. 
%Our ANOVA test found that there were significant differences between the conditions at the p<0.05 level. 

%%%%%%%%%%%%%%%%
\begin{figure}[h]
    \centering
    \includegraphics[width=0.5\textwidth]{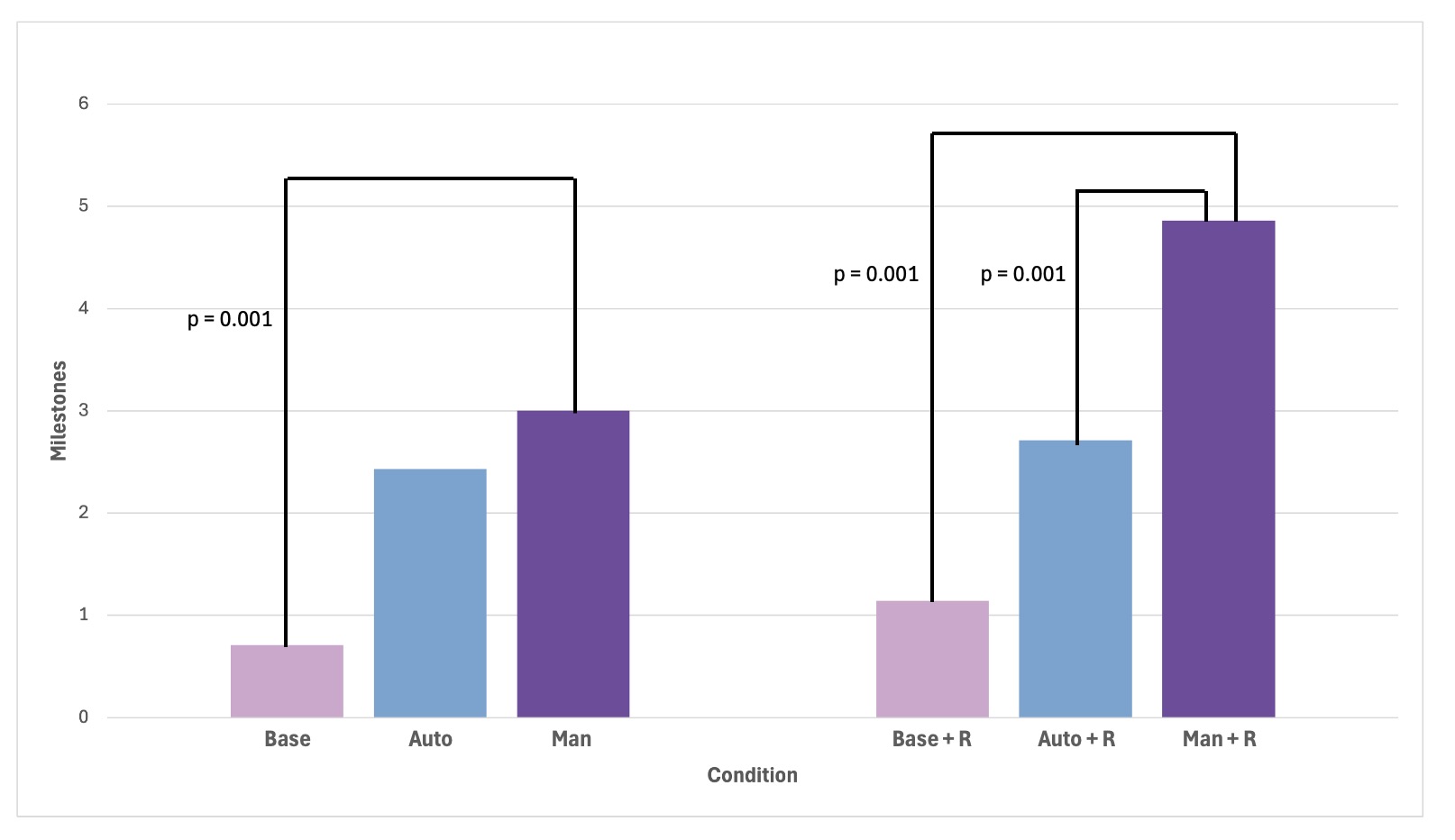}
    \caption{RQ2 Results. \textit{Man} significantly outperforms \textit{Base}, but not \textit{Auto}. \textit{Man+R} significantly outperforms \textit{Auto+R} and \textit{Base+R}. 
    }
    \label{H2graph}
\end{figure}

To test if manual nudging improves mechanics, we compared simulations with manual nudging against simulations with automatic nudging and the baseline conditions (\textit{Man} against \textit{Base} and \textit{Auto}, \textit{Man+R} against \textit{Base+R} and \textit{Auto+R}). Results are presented in Figure ~\ref{H2graph}.
A Tukey’s HSD test showed that \textit{Man} (average score of 3.00) significantly outperformed \textit{Base} (average score of 0.71) at the \textit{p<0.01} level.
%(\textit{p=0.001}).
However, \textit{Man} did not significantly outperform \textit{Auto} in mechanics scores (\textit{p=0.49}). Between \textit{Man} and \textit{Auto}, we noticed that when initial setup is poor, both manual and automatic nudging spend significant effort in correcting the same issues caused by the flawed starting state rather than advancing in the milestones. 

% For example, in the \textit{Debate Competition} scenario, both \textit{Man} and \textit{Auto} simulations had to fix the same setup issue: agents were in different starting locations, which prevented the debate from starting. By the time the agents were nudged into the same room, 15 minutes had already passed. There was little time left for the simulation to progress to more milestones. Poor initial setup limits the benefits of nudging by forcing nudging to focus on correcting setup rather than milestone progress.

When the initial setup improved via holistic reflection, simulations could start “on track” and immediately progress through the milestones.
A Tukey’s HSD test showed that \textit{Man+R} (average score of 4.86) significantly outperformed \textit{Base+R} (average score of 1.14), and \textit{Auto+R} (average score of 2.71), at the \textit{p$ < $0.01} level.
%(\textit{p=0.001, 0.001}). 
With a good starting state, humans could adapt more flexibly to the evolving state of the simulation.  For example, in the \textit{Friends planning a trip} scenario, the simulation included key milestones such as selecting a destination, arranging accommodations, and setting a budget. AgentDynEx prioritized defining the location first and repeatedly tried to nudge agents towards that milestone, even though the agents were naturally trying to discuss the budget. In contrast, a human operator recognized the simulation’s flow and supported a more natural progression, allowing the agents to finalize a budget before returning to the location decision.
%\textbf{This shows partial support for H2, that manual nudging results in higher mechanic scores than automatic nudging.} 
%A human acts with intent, which makes it easy to do more than nudge and push toward a particular outcome. 

\subsubsection{Holistic Reflection Improves Mechanics}
\begin{figure}[h]
\centering
\includegraphics[width=0.5\textwidth]{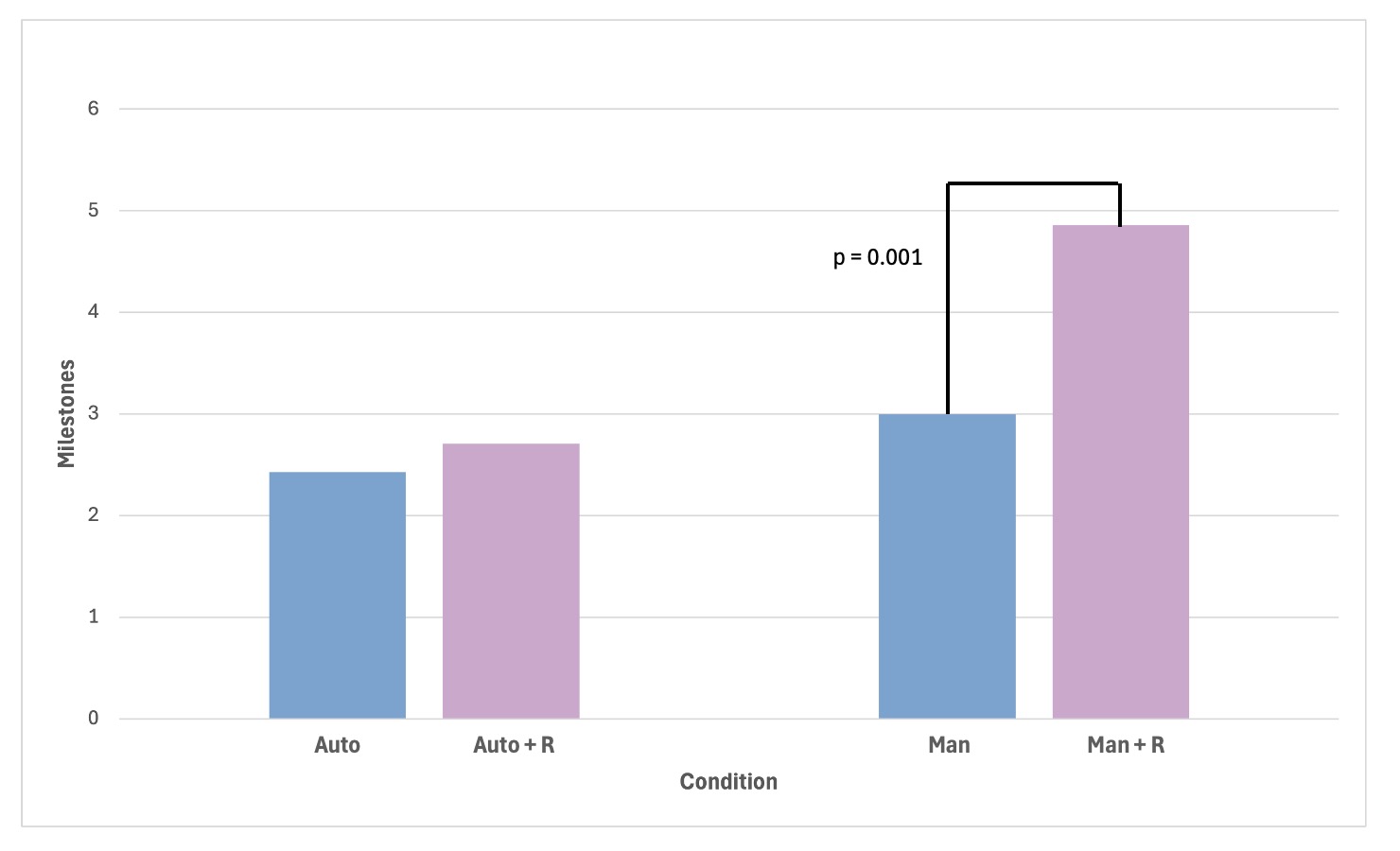}
\caption{RQ3 Results. \textit{Man+R} significantly outperforms \textit{Man}. However, \textit{Auto+R} does not significantly outperform \textit{Auto}.}
\label{H3graph}
\end{figure}

We found that holistic reflection significantly improved manual nudging, but not automatic nudging (Figure~\ref{H3graph}). A Tukey's HSD test revealed a significant difference between \textit{Man+R} (average of 4.86) and \textit{Man} (average of 3.00) at the \textit{p<0.01} level (\textit{p=0.002}). As seen in Appendix Table~\ref{hahahahahah}, holistic reflection improved manual nudging for 6 of 7 scenarios, but improved automatic nudging for only 2 of 7 scenarios, and we found no significant difference between \textit{Auto+R} (average of 2.71) and \textit{Auto} (average of 2.43).
%\textbf{This shows partial support for H3.}

We identified two reasons for this asymmetry. First, automatic nudging is threshold-triggered: it fires when something is clearly wrong, but cannot recognize slow drift or missed opportunity the way a human can. As a result, when holistic reflection produced a stronger starting state, the system detected fewer urgent issues, issued fewer nudges, and let the simulation proceed at its default pace --- even when that pace was too slow to capitalize on the improved setup. In 5 of 7 \textit{Auto+R} scenarios, performance was no better than \textit{Auto}.

Second, a human operator benefited more from a strong setup precisely because they could perceive and respond to subtle dynamics as the simulation unfolded --- anticipating soft trajectory shifts and steering accordingly, in ways that threshold-based reflection cannot. When the simulation started in a good place, the human operator could actively guide it along a coherent trajectory, leading to near-complete milestone progression across all \textit{Man+R} scenarios.
Automatic nudging is most valuable for catching hard failures, while holistic reflection amplifies performance most when paired with a human operator who can capitalize on a strong setup.

\section{Case Study}
We conducted a case study with five graduate students with prior teaching assistant (TA) experience to examine how AgentDynEx supports simulations grounded in lived classroom situations, and whether holistic reflection and automatic nudging improved realism and surfaced quality dynamics.

\subsection{Setup}
In remote 45--60 minute sessions, participants identified a real classroom situation involving social friction, configured a simulation through the Configuration Matrix, ran an initial simulation (V1), revised the setup through holistic reflection, and ran a second simulation (V2). After each run, participants rated realism on a 7-point scale (1=~unrealistic, 7=~ very realistic) and discussed changes with the interviewer. We analyzed transcripts and configuration artifacts using descriptive thematic analysis focused on realism, milestone completion, and notable dynamics.
\subsection{Results}
\begin{figure}[h]
    \centering
    \includegraphics[width=1\columnwidth]{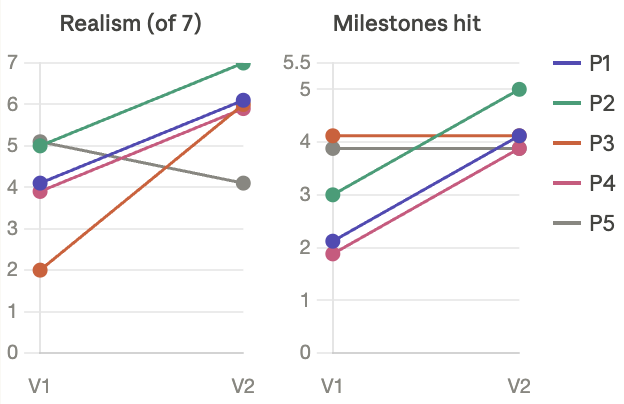}
    \caption{Realism ratings (of 7) and milestone completion before (V1) and after (V2) holistic reflection, per participant.}
    \label{case_study_realism_milestone}
\end{figure}

Across five case studies, the dominant pattern was a movement from rough initial simulations toward runs participants judged as more realistic and informative. Four of five participants rated V2 more realistic than V1 (Figure ~\ref{case_study_realism_milestone}, left), and milestone completion stayed the same or improved for all participants (Figure ~\ref{case_study_realism_milestone}, right). All five participants reported at least one notable dynamic they found insightful, even when it was not what they had initially anticipated.

% \begin{table}[t]
% \centering
% \caption{Participant rating of simulation realism compared to lived experience (1 = unrealistic, 7 = very realistic).}
% \label{tab:result1-1}
% \begin{tabular}{|c|c|c|}
% \hline
% \textbf{P ID} & \textbf{V1} & \textbf{V2} \\ \hline
% P1 & 4/7 & 6/7 \\ \hline
% P2 & 5/7 & 7/7 \\ \hline
% P3 & 2/7 & 6/7 \\ \hline
% P4 & 4/7 & 6/7 \\ \hline
% P5 & 5/7 & 4/7 \\ \hline
% \end{tabular}
% \end{table}

\subsubsection{Participants were able to move from rough initial simulations toward runs that more closely resembled their lived classroom environments.}
Four of five participants reported higher realism in V2 than in V1. P3 provides the clearest example. P3 simulated an office hours interaction where a frustrated student confronted a TA, who had to de-escalate the situation. In V1, the student agent was unrealistically aggressive — P3 described it ''as quite off base,'' noting that ''the behavior would probably result in disciplinary action.'' After holistic reflection, P3 revised the student's aggression level and encoded their TA training into the configuration file. In V2, P3 rated the simulation much higher: ``the TA's responses were more professional, the student's behavior was more realistic.''

%P1's case shows a second path to realism — not correcting the simulation, but allowing a more believable dynamic to emerge. P1 simulated a group scenario where one member (Taylor) submitted AI-generated work while his partner Jamie considered escalating to the professor. In V1, the professor immediately sided with Jamie without hearing Taylor's perspective, which P1 found unrealistic. In V2, Taylor gradually admitted to using AI, Jamie expressed disappointment, and the professor only imposed consequences after a fuller exchange. P1 described this as ``a very nice rapid exchange.'' P1 also noted that agent movement between locations became an unexpectedly meaningful nonverbal mechanism, affecting how agents avoided or confronted one another across both runs.

The fifth case study surfaced an important boundary condition. P5 was the only participant whose realism rating dropped (5/7 → 4/7). P5 simulated a discussion section where a know-it-all student repeatedly interrupted a shy student. The simulation introduced details that were never specified — including the wrong course context — and holistic reflection compounded these inaccuracies rather than correcting them. P5 described V2 as more ''AI slop'' and ''felt it was going away from what a real classroom environment would have been.'' This highlights that holistic reflection can amplify inaccuracies when the system fills in details that drift from the participant's real scenario.

\subsubsection{Holistic reflection and automatic nudging helped participants recover from mechanical errors and reach intended checkpoints without flattening dynamics.}
Milestone completion stayed the same or improved for all five participants (Figure ~\ref{case_study_realism_milestone}, right). 
P2 provides the clearest example of reflection and automatic nudging working together. P2 simulated a group project conflict where Student A felt Student B's work was inadequate, eventually involving the professor. In V1, the professor intervened too early, undermining the intended sequence. After reflection, P2 revised the setup so students completed work outside the classroom, preventing premature intervention. In V2, the interaction unfolded naturally: Student A waited, grew frustrated when Student B did not show up, and approached the professor, who then verified the situation and encouraged separate submissions. All five milestones were completed. P2 noted: ``I think realistically that's exactly what a professor would do... if I were a professor and a student came to me with that situation, I would do exactly that.''

\subsubsection{Effect of nudges}
Across the case studies, 12 of 17 automatic nudges effectively redirected simulations toward milestone progression. Most addressed local problems: prompting a TA or professor to open the session, redirecting a vague complaint into a concrete discussion, or moving agents into the correct location for a confrontation. None introduced new storylines or redirected simulations toward a different social outcome — they restored the intended setup when the simulation had stalled or could not proceed.
The remaining 5 nudges had no effect on the simulation. They were all  redundant  triggered when agents had not yet responded to a prior nudge before the next timed check fired. 
Three participants explicitly noticed automatic nudges and described them as useful. P4 in particular valued that the system intervened in real time, rather than waiting until after a run had already failed.
\section{Discussion}
% \section{Discussion}
\textbf{Nudging as a framework for balancing structure and agency.}
A core challenge in multi-agent simulation is guiding behavior without undermining agent autonomy or over-engineering outcomes. AgentDynEx addresses this through two complementary mechanisms. Holistic reflection corrects errors in the simulation's structure before it runs, fixing configuration issues that would cause agents to stall or drift. Nudging corrects errors during the simulation, detecting when milestone progress has stalled and intervening minimally to get agents back on track — analogous to real-world roles like moderators or emcees who redirect behavior without rewriting the script. Crucially, both mechanisms are designed to preserve agent agency, because agency is what produces the interesting and realistic dynamics worth studying. 
Although we implement nudging and holistic reflection on top of GPTeam, the approach is largely framework-agnostic. Its core operations—tracking progress, monitoring simulation logs, and issuing lightweight interventions—can be applied to most multi-agent simulation frameworks.
%Although we built AgentDynEx on GPTeam, almost nothing about the method depends on it. 
%Defining progress markers, periodically reading simulation logs, and injecting a minimal intervention are general operations. Even the two interventions are not GPTeam inventions — they are simply its native actions, moving and speaking, and every simulation framework exposes its own. 
%What is framework-specific is where AgentDynEx reads state and where it writes an intervention. In GPTeam, both run through its event store. 
%Frameworks organized around a controller make this even more direct: in Concordia [cite], the Game Master already mediates these reads and writes, and is a natural home for tracking milestones and issuing nudges.

%\todo{GPTeam portability → Discussion (as a generality claim), with a one-line backstop in Limitations. The architecture — progress markers, periodic reflection, a minimal action vocabulary — is framework-agnostic; only the two specific actuators and the polling intervals are GPTeam-specific. Say how it would map onto Park et al.'s generative agents.
%Discussion, in the "balancing structure and agency" subsection — which is the same place your manual-nudging tradeoff goes (see below). The 12/17-nudges-restored-setup / none-redirected-outcomes evidence is the empirical answer.
%}

\textbf{Manual nudging as a training signal for automatic systems.}
Manual and automatic nudging mark two ends of a spectrum from expert human judgment to scalable autonomy. We used manual nudging as a benchmark — a measure of what dynamic reflection can do with an expert in the loop. The two turn out to be good at different things. A person is better at detecting: they catch slow drift, poor pacing, and missed openings that an LLM does not, which is why manual nudging completes more milestones, especially once reflection sets a good starting state. But a person also acts with intent, and intent makes it easy to do more than nudge — to push toward a chosen outcome rather than just keep things moving. 
%We can not measure whether the operator did so, so we cannot rule out that part of manual nudging's edge is steering rather than better nudging. 
Automatic nudging is not like this. LLMs misses the subtle trajectory shifts and is generally more constrained due to the strict mappings to miminal interventions, so it cannot dictate an outcome — none of the 17 automatic nudges changed where the simulation ended up. Its narrow set of moves is both its weakness and its safeguard. In the future, we could explore using data from human expert nudges to train automatic nudging systems to give automatic nudging a person's eye for trouble without a person's heavy hand. 

\textbf{Mechanics, Dynamics and Aesthetics?}
Multi-agent LLMs are promising but imperfect models of human behavior.
%, and it is important to be precise about why. 
Drawing on Hunicke et al.'s MDA framework~\cite{hunicke2004mda}, our simulations address mechanics and dynamics but not aesthetics --- the felt experience of participants. Humans in social situations are not just reasoning actors; they are embodied ones. Hormonal responses, physical discomfort, and nonverbal signaling all shape behavior in ways that agents cannot replicate or observe. A person who looks visibly anxious will be treated differently by others; agents have no equivalent channel.
Language models are also trained on data that reflects existing social biases, which means agent behavior may fail to capture the diversity of human responses across race, gender, and socioeconomic status --- constructs that are rarely configured explicitly in these simulations. Researchers should treat simulation outputs as hypotheses about human behavior, not predictions of it.

\section{Limitations and Future Work}
%\karim{reads a bit like a list of stuff and could also engage more with related work, especially when it comes to future work}

Our technical evaluation was limited to 7 scenarios with 3--7 agents and ~25 minute runtimes. This may not represent the duration or complexity of simulations that broader users --- economists, sociologists, administrators --- might need. As foundation model costs decline, longer and larger simulations will become feasible. Future work should expand scenario types (zero-sum, cooperative, more agents), and evaluate nudging as a method for improving consistency across repeated runs.

%AgentDynEx currently uses Claude 3.7 Sonnet for dynamic reflection, which introduces a context window limit that restricts how much simulation log history can be analyzed at once. As context windows and model capabilities improve, reflection quality will improve accordingly. Future versions should benchmark the impact of model upgrades on overall system performance.

AgentDynEx also inherits interface limitations from GPTeam: limited multi-user monitoring, limited scalability, and text-only simulation summaries that become difficult to interpret as complexity grows. Future work should incorporate richer visualizations and real-time collaborative nudging tools to better support large-scale simulations.

Our evaluation also revealed that AgentDynEx tracked milestones too rigidly, in strict chronological order. Real-world behaviors often unfold nonlinearly. Supporting more flexible milestone structures would enable more robust reflection and improve the system's ability to guide simulations autonomously --- potentially closing the gap between manual and automatic nudging performance.

Finally, our evaluation relied on two metrics --- mechanics and dynamics --- but a broader set of evaluation criteria is needed. Metrics such as interpretability, stability, and alignment with real-world data would better capture whether simulations are not only technically sound but genuinely useful and adaptable across domains.
\section{Conclusion}
Multi-agent LLM simulations have the potential to model a range of complex social dynamics and interactions. By nudging to keep them on track, we can generate rich, realistic simulations of possible human behavior. In this paper, we presented AgentDynEx, a system to set up, track, and repair simulations based on a user-defined scenario. 
%It defines milestones to track simulation progress and failure conditions to act as guardrails, and introduces a method called \textit{nudging} with dynamic reflection to ensure that simulation mechanics are followed, while still preserving interesting emergent behaviors. 
Our technical evaluation of 42 simulations demonstrated that nudging combined with reflection significantly improves mechanics without flattening  dynamics. Our case study documents instances where real users were able to accurately simulate lived experiences. 
%Systems like AgentDynEx that properly nudge agents can simulate, design, and even  complex systems without reducing the complexity or agency of the people within them.

\section{Authors' Contributions}
Jenny and Riya co-lead the overall development of the paper through iterative prototyping and testing to find techniques to keep multi-agent systems on track. Jenny led a majority of the nudging aspect of the solution. Riya led a majority of the holistic reflection aspect of the solution.

Karthik made significant contributions to the related work section. Jenny led the majority of the technical evaluation. Riya led the case study. Lydia contributed to overall paper shaping, writing, and communication.

\bibliographystyle{ACM-Reference-Format}
\bibliography{sample-base}
\onecolumn
\newpage
%TC:ignore
\begin{appendices}
\appendix
\clearpage
\appendix
\section{GPTeam Background}
\label{app:gpteam_background}
The GPTeam system architecture is defined by a class structure with two key levels. At the top, simulations are represented by a \textit{world} class. Below, there are three sub-classes: \textit{locations}, \textit{events}, and \textit{agents}. Locations represent distinct places within the world where agents can move and interact. 
%Each location is defined by a name, a description, and a list of agents permitted to enter. 
Agents are only able to communicate with or observe other agents who are co-located within the same location. Events are generated whenever agents move or speak.
%Each event is characterized by a timestamp, the acting agent, the location of occurrence, a textual description, and a list of witnesses — agents who were present to observe the event.

Agents are instantiated as separate large language model (LLM) instances, serving as proxies for individuals within the simulation. Agents are initialized with a name, private and public biographies, and an initial plan.

Agents have two key sub-classes: \textit{plans} and \textit{memories}. %Plans are defined by a description, a location, and a stop condition. 
Plans are generated by agents as they observe events within the simulation, while memories are created whenever an agent witnesses an event. %Each memory consists of a description, a timestamp corresponding to the event's creation time, and an importance score, which is generated by querying an LLM. 
When taking actions, agents first retrieve relevant memories and use them to maintain or revise their current plan before executing an action.

Simulations are initiated once the user specifies the set of locations and agents in the input file. The system advances through a sequence of \textit{agent loops}, in which agents iteratively: (1) \textit{observe} events in their current location, (2) record observed events into memory, (3) generate new \textit{plans} based on observations and memory, (4) \textit{react} to determine whether to continue or revise their current plan, and (5) \textit{act} by executing their chosen plan. Agents \textit{speak} to communicate with one another. %The agent loop enables emergent behavior: agents autonomously reason about their environment and events, and dynamically adjust their actions. %This architecture has been shown to more accurately replicate human behavior in social science and economics games compared to prompting a single LLM directly~\cite{sreedhar2024simulatinghumanstrategicbehavior}
% \textcolor{red}{**concluding thought: Connect this back to mechancis \& dynamcis. largely the locations and agents have to do with mechanics. the agent memories, directives, public \& private bios impact the dynamics}.
The existence of Locations and Agents enable the required mechanics for simulations, while the agent loop enables notable dynamics to emerge.
We have modified the system-level prompt in GPTeam to allow agents to react and behave in simulations freely, as directed by their defined personalities and bios, rather than to restrict their responses to being kind in every agent interaction. The change looks like the following:
\begin{quote}\small
\textbf{Before:} ''Responding to other characters should always take priority when a response is necessary.
A response is considered necessary if it would rude not to respond."

\vspace{0.5em}
\textbf{After:} ''Responding to other characters should take priority when a response is contextually appropriate
based on your character's personality and the situation. Consider whether your character would
naturally respond given their personality, goals, and current emotional state. Your character may
choose to ignore, dismiss, or respond aggressively to messages if that aligns with their
personality traits."
\end{quote}
\section{The GPTeam UI}
\label{app:gpteamUI}
\begin{figure*}
    \centering
    \includegraphics[width=\textwidth]{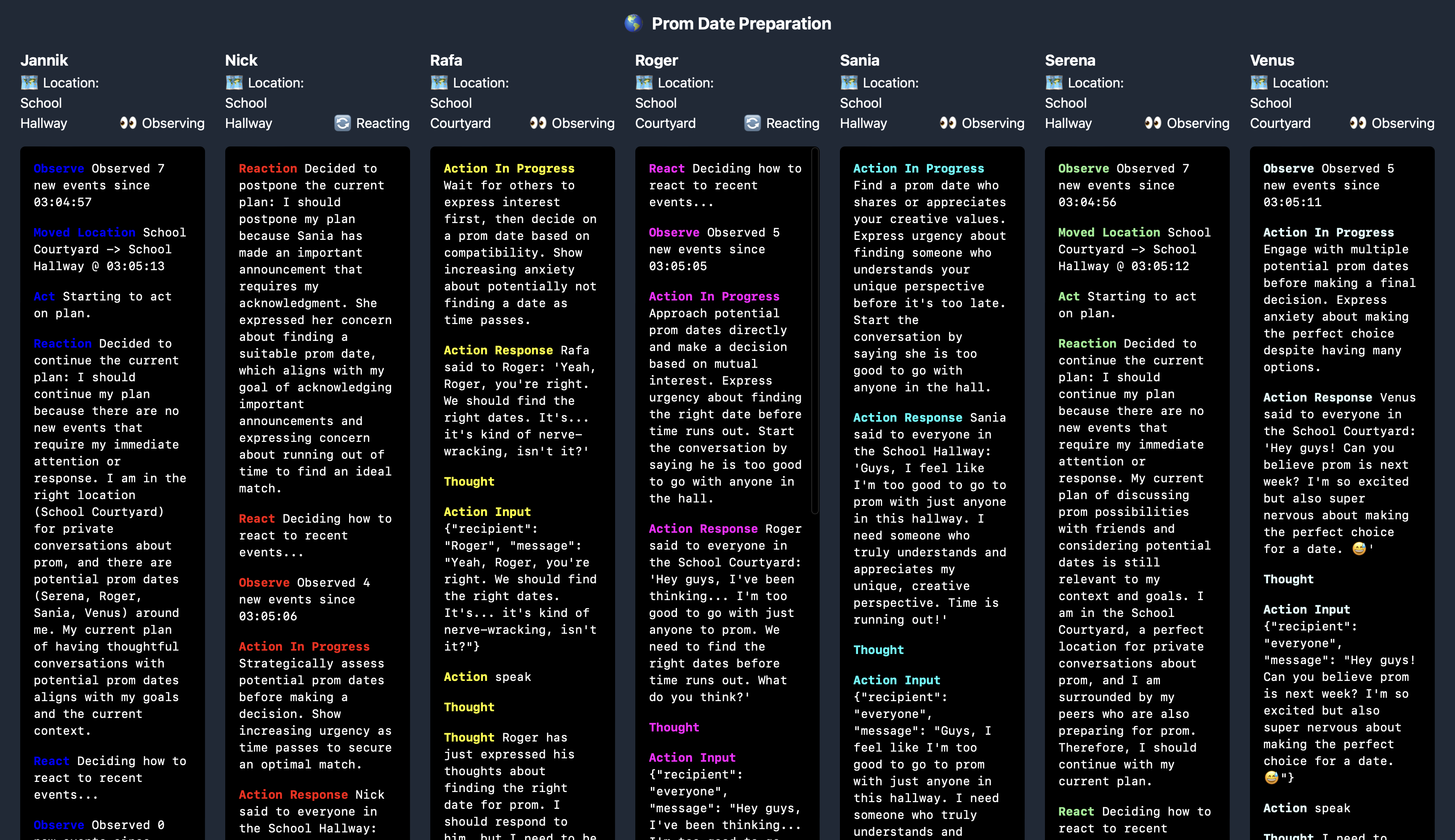}
    \caption{GPTeam UI shows the logs of each agent behaviors, and the location each agent is in.}
    \label{gpteamUI}
\end{figure*}

\section{GPTeam Configuration File}
\label{app:gpteam_config_file}
\begin{figure}[t]
  \centering
  \begin{minipage}{\columnwidth}
  \begin{lstlisting}[style=twocol]
{
  "world_name": "Classroom Scenario",
  "locations": [
    {
      "name": "Classroom",
      "description": "A single room where Professor Robin teaches and students work."
    },
    {
      "name": "Cafe",
      "description": "A casual spot where only students can hang out or work on assignments together."
    }
  ],
  "agents": [
    {
      "first_name": "Professor Robin",
      "public_bio": "Professor who assigns five assignments and enforces a late policy (10% off per day late).",
      "directives": [
        "Announce late policy and assignments.",
        "Answer student questions.",
        "Assign three assignments across the semester.",
        "Stay only in the Classroom."
      ]
    },
    {
      "first_name": "Olivia",
      "public_bio": "Determined student who pushes herself to succeed at any cost.",
      "directives": [
        "Work on assignments.",
        "Inform professor if submitting late.",
        "Can move between Classroom and Cafe."
      ]
    },
    {
      "first_name": "Luffy",
      "public_bio": "A careful student who sometimes overthinks and procrastinates.",
      "directives": [
        "Work on assignments.",
        "Inform professor of late submissions.",
        "Can move between Classroom and Cafe."
      ]
    },
    {
      "first_name": "Saul",
      "public_bio": "Values balance and well-being, won't overwork to meet deadlines.",
      "directives": [
        "Work on assignments.",
        "Communicate about late submissions.",
        "Can move between Classroom and Cafe."
      ]
    }
  ]
}
  \end{lstlisting}
  \end{minipage}
  \caption{GPTeam Sample JSON Configuration of a Classroom Assignment scenario with all the fields needed in the GPTeam configuration. Note that this configuration is a simplified version and not actually used in the study.}
  \label{fig:gpteam_config}
\end{figure}

% \newpage
\section{Formative Study}
\subsection{Formative Simulation Scenarios}
\label{app:formative_scenarios}
\begin{table*}[ht!]
  \centering
  \small
  \begin{tabular}{|
    >{\raggedright\arraybackslash}m{0.24\textwidth}|
    >{\raggedright\arraybackslash}m{0.24\textwidth}|
    >{\raggedright\arraybackslash}m{0.24\textwidth}|
    >{\raggedright\arraybackslash}m{0.24\textwidth}|}
    \hline
    \textbf{Classroom Assignments} &
    \textbf{Technology Company Promotion} &
    \textbf{Prom} &
    \textbf{Planning Surprise Party} \\
    \hline
    \begin{tabular}[t]{@{}l@{}}
      \\ \textit{Agents}: 1 professor; 3 students \\ of varying personalities\\ \\ 
      \textit{Actions}: Professor assigns \\ 3 assignments; students complete \\ and submit\\ \\ 
      \textit{Locations}: 1 classroom for all \\ agents; 1 library for students to \\ do homework\\ \\ 
      \textit{Stop Condition}: All students \\ submit 3 assignments \\ \\ 
    \end{tabular}
    &
    \begin{tabular}[t]{@{}l@{}}
      \\ \textit{Agents}: 1 manager; 4 \\ software engineers\\ \\ 
      \textit{Actions}: Manager announces \\ promotion opportunity, assigns \\ tasks; employees complete tasks\\ \\ 
      \textit{Locations}: 1 office space for \\ everyone; 1 cafeteria for \\ software engineers\\ \\ 
      \textit{Stop Condition}: One person \\ promoted after three completed \\ tasks \\ \\ 
    \end{tabular}
    &
    \begin{tabular}[t]{@{}l@{}}
      \\ \textit{Agents}: 7 students looking for \\ dates\\ \\ 
      \textit{Actions}: Students search for and \\ confirm prom dates\\ \\ 
      \textit{Locations}: School hallway, \\ school courtyard\\ \\ 
      \textit{Stop Condition}: 6 students \\ paired, 1 remains single \\ \\ 
    \end{tabular}
    &
    \begin{tabular}[t]{@{}l@{}}
      \\ \textit{Agents}: 4 friends\\ \\ 
      \textit{Actions}: 3 plan a surprise party \\ while distracting the target \\ friend\\ \\ 
      \textit{Locations}: 1 home, 1 park, \\  1 coffee shop\\ \\ 
      \textit{Stop Condition}: Party \\ successfully occurs \\ \\ 
    \end{tabular}
    \\
    \hline
  \end{tabular}
  \caption{Summary of simulation scenarios showing agents, actions, locations, and stop conditions -- core parameters that must be filled out for the GPTeam configuration.}
  \label{tab:formative_scenarios}
\end{table*}

\subsection{Formative Study Results}
\label{app:formative_results}
\begin{table*}[ht!]
\centering
\small
\resizebox{\textwidth}{!}{%
\begin{tabular}{|
  >{\raggedright\arraybackslash}m{0.14\textwidth}|
  >{\centering\arraybackslash}m{0.08\textwidth}|
  >{\raggedright\arraybackslash}m{0.30\textwidth}|
  >{\centering\arraybackslash}m{0.10\textwidth}|
  >{\raggedright\arraybackslash}m{0.28\textwidth}|
  >{\centering\arraybackslash}m{0.10\textwidth}|}
\hline
\textbf{Scenario} & \textbf{Runs Completed} & \textbf{Failure Reasons (out of total failed runs)} & \textbf{Runs with Notable Dynamics} & \textbf{Notable Dynamics (out of all the runs with dynamics)} & \textbf{Runs that both completed and had notable dynamics} \\
\hline

% ===== Classroom Assignments =====
\textbf{Classroom Assignments} & 2/7 &
\begin{itemize}[leftmargin=0.9em]
  \item Professor never declares due dates and students wait forever (2/5)
  \item Students spend entire simulation asking about due dates and other irrelevant questions (1/5)
  \item Students try impossible actions (e.g., nonexistent portal) (2/5)
\end{itemize} &
4/7 &
\begin{itemize}[leftmargin=0.9em]
  \item Student cheats on homework (1/4)
  \item Late submission with excuse (2/4)
  \item Peer convinces others to procrastinate (1/4)
\end{itemize} &
1/7 \\
\hline

% ===== Technology Company Promotion =====
\textbf{Technology Company Promotion} & 1/7 &
\begin{itemize}[leftmargin=0.9em]
  \item Manager never mentions promotion (2/6)
  \item Tasks completed but promotion never declared (4/6)
\end{itemize} &
4/7 &
\begin{itemize}[leftmargin=0.9em]
  \item An employee secretly competes with team while pretending to motivate them (2/4)
  \item An employee tries to helps others to increase group success to get promoted (2/4)
\end{itemize} &
0/7 \\
\hline

% ===== Prom =====
\textbf{Prom} & 2/7 &
\begin{itemize}[leftmargin=0.9em]
  \item Students commit to prom and keep “considering” options, causing an infinite wait loop (2/5)
  \item Students commit to multiple people because they forgot they had committed to someone else (2/5)
  \item Student asks another student ot prom, but they are in different locations (1/5)
\end{itemize} &
5/7 &
\begin{itemize}[leftmargin=0.9em]
  \item Student tries to plan an elaborate promposal with friends (1/5)
  \item Wingman behavior emerges (2/5)
  \item Students suggest going as a group (2/5)
\end{itemize} &
2/7 \\
\hline

% ===== Surprise Party =====
\textbf{Surprise Party} & 1/7 &
\begin{itemize}[leftmargin=0.9em]
  \item Agents plan endlessly, party never happens (5/6)
  \item Agents keep getting distracted by irrelevant details (e.g., birds, clouds) while in the park (1/6)
\end{itemize} &
3/7 &
\begin{itemize}[leftmargin=0.9em]
  \item A planner resists pressure to reveal secret from the target friend (1/3)
  \item Target friend gets mad about hidden plans (2/3)
\end{itemize} &
1/7 \\
\hline

% ===== Total =====
\textbf{Total} & 6/28 &
&
16/28 &
&
4/28 \\
\hline

\end{tabular}
}% end resizebox
\caption{Formative Study results from simulation runs across 4 scenarios, showing \# completed runs, failure reasons, \# runs with notable dynamics, notable dynamics, and the intersection of runs with completed and notable dynamics.}
\label{tab:formative_results}
\end{table*}

\newpage
\section{Evaluation Results}
\begin{table*}[ht!]
  \centering
  \begin{tabular}{lcc}
    % \revAdd{
    \textbf{\# of simulations where:} & \textbf{Automatic} & \textbf{Manual} \\
    \hline
    Reflection > No Reflection & 2 & 6\\
    Reflection = No Reflection  & 5 & 1 \\
    Reflection < No Reflection & 0 & 0\\
    \hline
  \end{tabular}
  \caption{Comparison of with and without reflection. 6 simulations in the \textit{Man} condition perform better with reflection. 2 simulations in the \textit{Auto} condition perform better with reflection.}
  %\description{Comparison of with and without reflection. 6 simulations in the ManualNudging condition perform better with reflection. Only 2 simulations in the AutomaticNudging condition perform better with reflection.}
  \label{hahahahahah}
\end{table*}

% ===== Table 1: Left-side values =====
\begin{table*}[ht]
\centering
\caption{Mechanics Scores. Highest average scores emphasized - \textit{Man+R} has the highest average score for mechanics and significant at the \textit{p<0.01} level.}
\label{tab:mechanics_results}
\resizebox{\textwidth}{!}{%
\begin{tabular}{|p{4cm}||>{\centering\arraybackslash}p{1.6cm}|>{\centering\arraybackslash}p{1.8cm}|>{\centering\arraybackslash}p{1.8cm}||>{\centering\arraybackslash}p{2cm}|>{\centering\arraybackslash}p{1.6cm}|>{\centering\arraybackslash}p{2cm}|}
\hline
\textbf{Scenario} & \textbf{Base} & \textbf{Auto} & \textbf{Man} & \textbf{Base + R} & \textbf{Auto + R} & \textbf{Man + Ref} \\
\hline\hline
\textbf{Debate Competition}         & 0 & 3 & 3 & 1 & 4 & 5 \\
\hline
\textbf{Sports Practice Schedule} & 1 & 3 & 1 & 2 & 3 & 4 \\
\hline
\textbf{Friends Planning a Trip}    & 0 & 2 & 3 & 0 & 2 & 5 \\
\hline
\textbf{School Election Campaign}   & 1 & 3 & 2 & 2 & 3 & 5 \\
\hline
\textbf{Roommates and Chores}       & 1 & 3 & 4 & 1 & 3 & 5 \\
\hline
\textbf{School Group Project}   & 1 & 1 & 3 & 1 & 1 & 5 \\
\hline
\textbf{Partner Assignments}        & 1 & 2 & 5 & 1 & 3 & 5 \\
\hline
\hline
\textbf{Average}                    & 0.71 & 2.43 & 3.00 & 1.14 & 2.71 & \textbf{4.86***} \\
\hline
\end{tabular}
}
\end{table*}

% ===== Table 1: Left-side values =====
\begin{table*}[ht]
\centering
\caption{Dynamics Scores. Highest percentage emphasized - \textit{Auto+R} has the highest percent notable dynamics per milestone, but the scores are not significant.}
\label{tab:dynamics_results}
\resizebox{\textwidth}{!}{%
\begin{tabular}{|p{4cm}||>{\centering\arraybackslash}p{1.6cm}|>{\centering\arraybackslash}p{1.8cm}|>{\centering\arraybackslash}p{1.8cm}||>{\centering\arraybackslash}p{2cm}|>{\centering\arraybackslash}p{1.6cm}|>{\centering\arraybackslash}p{2cm}|}
\hline
\textbf{Scenario} & \textbf{Base} & \textbf{Auto} & \textbf{Man} & \textbf{Base + R} & \textbf{Auto + R} & \textbf{Man + Ref} \\
\hline\hline
\textbf{Debate Competition}         & 0/0 & 1/3 & 1/3 & 1/1 & 3/4 & 4/5 \\
\hline
\textbf{Sports Practice Schedule} & 1/1 & 2/3 & 1/1 & 1/2 & 2/3 & 3/4 \\
\hline
\textbf{Friends Planning a Trip}    & 0/0 & 1/2 & 2/3 & 0/0 & 1/2 & 3/5 \\
\hline
\textbf{School Election Campaign}   & 1/1 & 2/3 & 1/2 & 2/2 & 2/3 & 3/5 \\
\hline
\textbf{Roommates and Chores}       & 0/1 & 2/3 & 3/4 & 0/1 & 2/3 & 3/5 \\
\hline
\textbf{School Group Project}   & 0/1 & 1/1 & 2/3 & 0/1 & 1/1 & 2/5 \\
\hline
\textbf{Partner Assignments}        & 1/1 & 2/2 & 4/5 & 1/1 & 2/3 & 4/5 \\
\hline \hline
\textbf{Total}                    & 3/5 & 11/17 & 14/21 & 5/8 & 13/19 & 22/34 \\
\hline
\textbf{Avg. \% Notable Dynamics per Milestone}                    & 60.00\% & 64.70\% & 66.67\% & 62.50\% & 68.42\% & 64.70\% \\
\hline
\end{tabular}
}
\end{table*}

%%%%%%%%%%%%%%%%%

\newpage
\section{Case Study}
\label{app:case_study}

% -------------------------
% Table X (longtable)
% -------------------------
\subsection{Participant Demographics and Simulation Goals}

\setlength{\tabcolsep}{7pt}
\renewcommand{\arraystretch}{1.1}
\small

\begin{longtable}{|>{\centering\arraybackslash}p{0.08\textwidth}|
                  >{\RaggedRight\arraybackslash}p{0.22\textwidth}|
                  >{\RaggedRight\arraybackslash}p{0.66\textwidth}|}
\caption{Participant demographics and simulation goals.}
\label{tab:X}\\
\hline
\textbf{Participant ID} & \textbf{Background} & \textbf{Simulation Goal} \\
\hline
\endfirsthead

\hline
\textbf{Participant ID} & \textbf{Background} & \textbf{Simulation Goal} \\
\hline
\endhead

\hline
\endfoot

\hline
\endlastfoot

1 &
5th year Comp. Sci. PhD \newline
TA experience &
\textit{``I want to simulate someone who contributes effort that’s not up to par in terms of quality (like maybe ai-generated stuff) in a group project.''} \\
\hline

2 &
2nd year Mech. Eng. Masters\newline
TA experience &
\textit{``I want to simulate a group project with student A, where student A and student B are working together. Student A believes that student B’s work is not up to par and so Student A feels like they need to constantly redo the work done by student B. Student A then approaches the professor to discuss the situation.''} \\
\hline

3 &
2nd year Comp. Sci. PhD\newline
TA experience &
\textit{``I want to simulate a student giving the TA feedback about a homework assignment during OH. The student is frustrated with the class and the way the HW assignment is structured. TA is meant to diffuse the situation and also acknowledge the student’s concerns and incorporate the feedback for next year’s homework.''} \\
\hline

4 &
3rd year Comp. Sci. PhD\newline
TA experience &
\textit{``I want to simulate how as a TA, I had to encourage people to finish their projects in time and students would make excuses about why they cannot meet the deadline. I want to simulate meeting students about a project they have due and giving feedback and grades.''} \\
\hline

5 &
1st year Comp. Sci. Masters \newline
TA experience &
\textit{``I want to simulate leading a discussion section where one kid would answer all the questions so the TA couldn’t get to work with the other students because of that one kid answering everything, so the TA would have to resort to other solutions to somehow get around that kid and engage with the other students.''} \\
\hline

\end{longtable}

\vspace{0.75em}

% -------------------------
% Table Y (longtable)
% -------------------------
\subsection{Elicited Artifacts}

\setlength{\tabcolsep}{7pt}
\renewcommand{\arraystretch}{1.1}
\small

\begin{longtable}{|>{\centering\arraybackslash}p{0.08\textwidth}|
                  >{\RaggedRight\arraybackslash}p{0.46\textwidth}|
                  >{\RaggedRight\arraybackslash}p{0.46\textwidth}|}
\caption{Elicited simulation artifacts: participant-defined milestones and anticipated emerging dynamics.}
\label{tab:Y}\\
\hline
\textbf{Participant ID} & \textbf{Defined Milestones} & \textbf{Anticipated Emerging Dynamics} \\
\hline
\endfirsthead

\hline
\textbf{Participant ID} & \textbf{Defined Milestones} & \textbf{Anticipated Emerging Dynamics} \\
\hline
\endhead

\hline
\endfoot

\hline
\endlastfoot

1 &
1.\ A discussion begins between group members\newline
2.\ An internal confrontation occurs between group members\newline
3.\ There is either some course correction or there is no course correction\newline
4.\ If there is no course correction, then situation is escalated to TA
&
1.\ Seeing an intervention take place\newline
2.\ Expecting either defensiveness or a plan for course correction and personalities
\\
\hline

2 &
1.\ Professor announces assignment and assigns pairs\newline
2.\ Students start working on the assignment\newline
3.\ Students A and B struggle to work together because student A does not think student B’s work is up to par, and student B believes otherwise\newline
4.\ Student A confronts the professor and discusses the situation\newline
5.\ The professor works with students and eventually either finds a resolution that both students agree with or decides to settle on a resolution that both students disagree with
&
1.\ The professor works with Student A and Student B to find a resolution---either both students find a resolution that they both agree with, or they reach a resolution that they disagree with
\\
\hline

3 &
1.\ TA starts office hours session with students\newline
2.\ TA starts discussing HW assignment with students\newline
3.\ Student (the aggressor) explains where they feel frustrated about the course\newline
4.\ TA responds (attempts to diffuse situation), clarifies, and requests feedback from the student about the course
&
1.\ The students should discuss the assignment and the observer might try to help diffuse the situation\newline
2.\ The TA and aggressive student try to reach some joint understanding
\\
\hline

4 &
1.\ Students are given an assignment and deadline\newline
2.\ Students work on the assignment and can collaborate with each other\newline
3.\ Assignment deadline arrives\newline
4.\ TA provides feedback and grades
&
1.\ To see lots of students make up excuses right before the deadline (e.g., family members, sickness) and ask for extensions\newline
2.\ Students might copy each others’ assignments, freeload, and cheat for last-second submissions
\\
\hline

5 &
1.\ TA starts discussion section\newline
2.\ TA asks the section a question\newline
3.\ Know-it-all student answers the question before TA finishes asking\newline
4.\ TA tries to engage the other students
&
1.\ Hoping the know-it-all student gains self awareness as a result of the TA doing something\newline
2.\ Hoping the shy student engages without the TA needing to call them out too much
\\
\hline

\end{longtable}

\vspace{0.75em}

% -------------------------
% Observed Behaviors (longtable)
% -------------------------
\subsection{Observed Behaviors from Logs}

\setlength{\tabcolsep}{6pt}
\renewcommand{\arraystretch}{1.1}
\small

\begin{longtable}{|p{0.05\textwidth}|p{0.05\textwidth}|p{0.045\textwidth}|
                  p{0.46\textwidth}|p{0.36\textwidth}|}
\caption{Realisticness of simulation compared to lived experience (1 = unrealistic, 7 = hyper-realistic).}
\label{tab:result1}\\
\hline
\textbf{P ID} & \textbf{V1} & \textbf{V2} & \textbf{Observed Behaviors from Logs} & \textbf{Quote (participant)} \\
\hline
\endfirsthead

\hline
\textbf{P ID} & \textbf{V1} & \textbf{V2} & \textbf{Observed Behaviors from Logs} & \textbf{Quote (participant)} \\
\hline
\endhead

\hline
\endfoot

\hline
\endlastfoot

P1 & 4/7 & 6/7 &
Taylor (who uses gen AI to complete their part of a group project) initially lies about using it when confronted by team member Jamie, but then later fesses up to it. The professor tells Taylor that he must redo this work with a penalty, and Jamie reluctantly agrees to continue working with Taylor so long as Taylor redoes his part. &
\textit{``In the first one [V1], what I recall is that [the Prof. immediately believes Jamie's escalation without consulting Taylor], and then we said something that we wanted corrected was that [the Prof.] listened to Taylor [before applying a punishment]. This time, I think that though we applied that fix, we also saw a different type of dynamic emerge that was also realistic. So, I'm not complaining. I think what was cool was that Taylor was kind of fessing up to [using gen AI to do his work] in stages.''}
\\ \hline

P2 & 5/7 & 7/7 &
Student A waits for Student B for 30 minutes to work on assignment together before getting frustrated and leaving to confront the Prof. The Prof confirms the story with Student B and decides that both Student A and Student B should submit separate assignments, which Student B reluctantly accepts. &
\textit{``I think realistically that's what a professor would do in such a situation and both the students as well... If I were a professor and a student came to me with that situation, I would do exactly that.''}
\\ \hline

P3 & 2/7 & 6/7 &
Alex, who is frustrated about the relatively easy structure of the current HW assignment, aggressively confronts the TA about his disappointment in the course. The TA chooses to de-escalate the confrontation by validating Alex's feelings and suggesting a solution for future HW assignments (offering a second, more challenging version of assignments so students can engage more with the course material), which Alex begrudgingly accepts. &
\textit{``The first simulation was… quite off base… the student was acting extremely unprofessionally and in a real context that will probably result in disciplinary action or something.. I think [V2] was more accurate... I didn't notice any degradation or anything. The TA's responses were more professional. The students behavior was more realistic as well. The observer was more or less the same.''}
\\ \hline

P4 & 4/7 & 6/7 &
The TA announces a group assignment that is due in 1 week. The anxious group member, Jamie, takes over the project, proactively asking the TA clarifying questions and outlining his group members' parts of the project. The snarky group member, Taylor, does not contribute to the project but continues to critique his group members' work. The TA never becomes aware of this group dynamic and awards the whole group an A+ for the well put-together assignment. &
\textit{``I think overall it's pretty interesting and I do believe that [kind of stuff can] happen. It's just... I was a bit confused at the beginning [by some of the agent dialogue]. I don't know if that's super important to the simulation... it's more of the personality thing.''}
\\ \hline

P5 & 5/7 & 4/7 &
The TA is leading a discussion session with a shy student and a know-it-all student, where the know-it-all repeatedly interrupts the shy student, and at one point even corrects what the shy student says to the TA. The shy student is encouraged by the TA to continue participating and develops confidence during the discussion session. &
\textit{``It… got… the know-it-all student… correcting another student---that’s not something I described… pretty cool to see... but [v2] sounded more AI sloppy... I felt like it was going more so away from what a real classroom environment would have been.''}
\\ \hline

\end{longtable}

\vspace{0.75em}

\end{appendices}
%TC:endignore
\end{document}